\documentclass{elsart}

\usepackage[square]{natbib}
\usepackage{units}

\usepackage{amssymb}

\usepackage[dvips]{graphicx}
\usepackage{psfrag}
\journal{Astroparticle Physics}

\begin{document}

\begin{frontmatter}



\title{Dependence of geosynchrotron radio emission on the energy and depth of maximum of cosmic ray showers}


\author[FZKIK]{T. Huege\corauthref{cor}},
\corauth[cor]{Corresponding author.}
\ead{tim.huege@ik.fzk.de}
\author[FZKIK]{R. Ulrich},
\author[FZKIK]{R. Engel}

\address[FZKIK]{Institut f\"ur Kernphysik, Forschungszentrum Karlsruhe, Postfach 3640, 76021 Karlsruhe, Germany}

\begin{abstract}
Based on CORSIKA and REAS2 simulations, we investigate the dependence of geosynchrotron radio emission from extensive air showers on the energy of the primary cosmic ray and the depth of the shower maximum. It is found that at a characteristic lateral distance, the amplitude of the bandpass-filtered radio signal is directly proportional to the energy deposited in the atmosphere by the electromagnetic cascade, with an RMS uncertainty due to shower-to-shower fluctuations of less than 3\%. In addition, the ratio of this radio amplitude and that at a larger lateral distance is directly related to the atmospheric depth of the shower maximum, with an RMS uncertainty of $\sim15$--20$\,$g$\,$cm$^{-2}$. By measuring these quantities, geosynchrotron radio emission from cosmic ray air showers can be used to infer the energy of the primary particle and the depth of the air shower maximum on a shower-to-shower basis.
\end{abstract}


\end{frontmatter}

% main text
%________________________________________________________________

\section{Introduction}

During the last few years, the technique of radio detection of cosmic ray air showers has experienced an impressive renaissance \citep{FalckeNature2005,HuegeCris2006,ArdouinBelletoileCharrier2005,vandenBergIcrc2007}. The activities are driven by the prospect of establishing a new observing technique with nearly 100\% duty cycle and very good angular resolution, which would complement the existing particle detectors and air fluorescence telescopes. Due to the coherent nature of the radio emission, i.e.\ an approximately quadratic scaling of the emitted radio power with the energy of the primary cosmic ray, the technique is particularly well-suited for the detection of ultra-high energy cosmic rays. 

One important question is how the observables measured with the radio technique can be related to the energy and mass of the primary cosmic ray particle. In this article, we use the geosynchrotron model for radio emission from cosmic ray air showers \citep{HuegeFalcke2003a,HuegeFalcke2005a,HuegeFalcke2005b,HuegeUlrichEngel2007a} to demonstrate how shower parameters could be derived from radio-only measurements that allow one to estimate the primary energy and composition. It is shown that the characteristics of the lateral distribution function of the radio emission can be exploited to estimate on a shower-to-shower basis both the energy deposited in the atmosphere and the depth of the shower maximum. The energy determination can be performed with a measurement at an optimum lateral distance which is independent of the energy of the primary particle in the considered energy range from $10^{18}$ to \unit[$10^{20}$]{eV} --- it only depends on the air shower geometry and radio observing frequency. Combining this measurement with an additional measurement at a different lateral distance provides a handle on the depth of the shower maximum and consequently yields information related to the mass of the primary particle. 

All results presented in this simulation study were derived with the geosynchrotron model as implemented in the REAS2 code \citep{HuegeUlrichEngel2007a}. Additional mechanisms which can contribute to the radio signal, such as Cherenkov-like emission from a charge excess and contributions from the net charge variation during the air shower evolution \citep{WernerScholten2008,MeyerVernetLecacheuxArdouin2008}, are the subject of current investigations and are not included in the analysis presented here. A direct comparison between the implementations of different emission models by various authors is difficult due to the very different calculational approaches. In contrast to other available models, however, the geosynchrotron model implemented in the REAS2 code allows us an absolute, parameter-free calculation of the emission from individual air showers, accounting for realistic shower-to-shower fluctuations and detailed spatial, angular and energy distributions of the shower particles. If radio emission from air showers is dominated by geomagnetic effects --- an assumption supported by experimental data showing strong correlations between the strength of the radio signal and the orientation of the air shower axis to the Earth's magnetic field \citep{FalckeNature2005,HornefferArena2005,PetrovicApelAsch2006} --- the results of this analysis can be considered generally applicable.

In the following we first describe the methodology applied throughout the present analysis. We then investigate the case of air showers with 60$^{\circ}$ zenith angle, where the effects are very prominent and could be well exploited experimentally. Afterwards we demonstrate that the same qualitative behaviour is fulfilled for air showers with 45$^{\circ}$ zenith angle, where it would, however, be harder to use experimentally, followed by our conclusions.

\section{Methodology}

We investigate four air shower geometries: showers with 60$^{\circ}$ zenith angle coming from the south (180+90 showers), from the east (180 showers) and from the north (180 showers) as well as air showers with 45$^{\circ}$ zenith angle coming from the south (180 showers). Each set of 180 showers covers the energy range from \unit[$10^{18}$]{eV} to \unit[$10^{20}$]{eV} according to a power-law with spectral index of $-1$ and consists of 60 proton-induced, 60 iron-induced and 60 photon-induced showers. The additional set of 90 showers has been simulated for a fixed energy of \unit[$10^{19}$]{eV}, induced in equal parts by proton, iron and photon primaries.

For each of these 810 showers, a complete simulation with CORSIKA 6.710 \citep{HeckKnappCapdevielle1998} using the hadronic interaction models QGSJETII-03 \citep{Ostapchenko2005,Ostapchenko2006b} and UrQMD 1.3.1 \citep{Bass1998,Bleicher1999} was performed to produce detailed histogram information to be used for the radio simulations as described in \citep{HuegeUlrichEngel2007a}. The magnetic field was set to a strength of \unit[0.23]{Gauss} with an inclination angle of -37.3$^{\circ}$, corresponding to the values valid for the southern site of the Pierre Auger Observatory \citep{AugerNIM2004} in Argentina. Particle thinning within CORSIKA was set to 10$^{-6}$ with optimised weight limitation according to \citep{Kobal2001}. Electromagnetic particles were tracked down to kinetic energies of \unit[400]{keV} using EGS4 \citep{EGS4}. For photon-primaries with energies $>10^{19}\,$eV, pre-showering in the Earth's magnetic field was taken into account. Also, the LPM effect \citep{LandauPomeranchuk1953a,LandauPomeranchuk1953b,Migdal1956} as implemented in CORSIKA is accounted for in the simulations.

In a second step, the radio emission for each of the 810 CORSIKA-simulated air showers was calculated with REAS 2.58 \citep{HuegeUlrichEngel2007a,REAS-Webpage} for a regular grid of observer locations at a height of \unit[1400]{m} above sea level, again adequate for the southern Pierre Auger Observatory. For the 45$^{\circ}$ and 60$^{\circ}$ zenith angle showers coming from the south, in addition, the radio signal at sea level was calculated.

The resulting unlimited bandwidth time-series pulses were then digitally filtered with three idealised rectangle filters: \unit[16--32]{MHz}, \unit[32--64]{MHz} and \unit[64--128]{MHz}. (The filtered pulses correspond to signals that experiments with the given frequency bandwidths would measure; the bandwidth of \unit[32--64]{MHz} is closest to the one employed in current experimental approaches.) The absolute values (vectorial sum of all three polarisation components) of the peak amplitudes of these filtered pulses, derived for each observer location and air shower, form the basis for the following discussion.

\section{The 60$^{\circ}$ zenith angle case}

We first discuss the important characteristics of the radio signal for the case of 60$^{\circ}$ zenith angle. In a first step we shall consider showers of \unit[$10^{19}$]{eV} energy and later show that the results also apply to the energy range from $10^{18}$ to \unit[$10^{20}$]{eV}.

\subsection{Characteristics of lateral profiles}

Earlier analyses \citep{HuegeFalcke2005b} have shown that the slope of the radio lateral distribution function is correlated with the depth of the air shower maximum, $X_{\mathrm{max}}$, and consequently contains information on the mass of the primary particle \citep{HuegeIcrc2005a}.

\begin{figure}[htb]
%\begin{minipage}{15.5pc}
\centering
\includegraphics[angle=270,width=20.0pc]{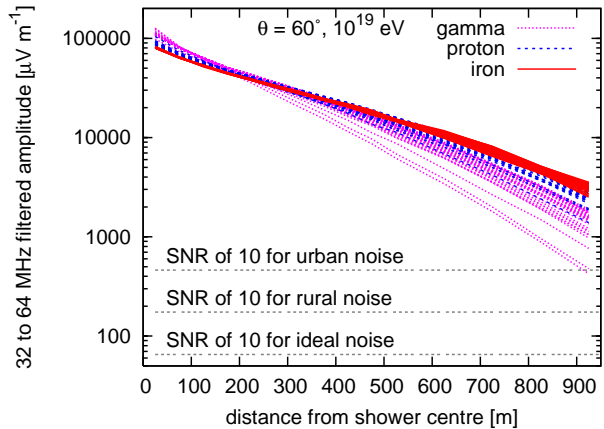}
\caption{\label{lateralslopes}Lateral distribution of the \unit[32--64]{MHz} filtered peak radio amplitude for \unit[$10^{19}$]{eV} showers coming from the south and observers north of the shower core. Estimates of peak radio amplitudes that would yield a signal-to-noise ratio (SNR) of 10 for ideal (galactic plus atmospheric), rural and urban noise as given in \citep{MeinkeGrundlach2007} are marked. Please note that the SNR scales with the filtered peak radio amplitude squared.}
%\end{minipage} \hspace{1.5pc}
%\begin{minipage}{15.5pc}
%\includegraphics[angle=270,width=15.5pc]{60deg0deg_1e19_lateralslopes_maxamp.eps}
%\caption{\label{bla}Bla.}
%\end{minipage}
\end{figure}

In Fig.\ \ref{lateralslopes}, we show the lateral distribution of the \unit[32--64]{MHz} filtered peak radio amplitude derived for the 90 air showers with fixed energy of \unit[$10^{19}$]{eV}. The lateral distance is given in ground coordinates\footnote{We use ground-distances rather than shower-distances throughout this article, as shower-distances would not remove the intrinsic asymmetries of the radio signal and ground distances are experimentally relevant in the end.} in the direction defined by the continuation of the air shower axis (i.e., in this case to the north, as the showers are coming from the south). Measurements at distances up to one km should be feasible for ultra-high energy cosmic rays, as the comparison with continuous noise estimates based on ITU/CCIR measurements taken from \citep{MeinkeGrundlach2007} illustrates.

There is an important feature in these lateral distributions: while the slopes of the individual lateral profiles differ significantly, all profiles intersect in a small region around $\sim 250\,$m. This essentially arises from the strong forward collimation of the radio emission in combination with the integration over the course of the air shower evolution.

\begin{figure}[htb]
\begin{minipage}{15.5pc}
\includegraphics[angle=270,width=15.5pc]{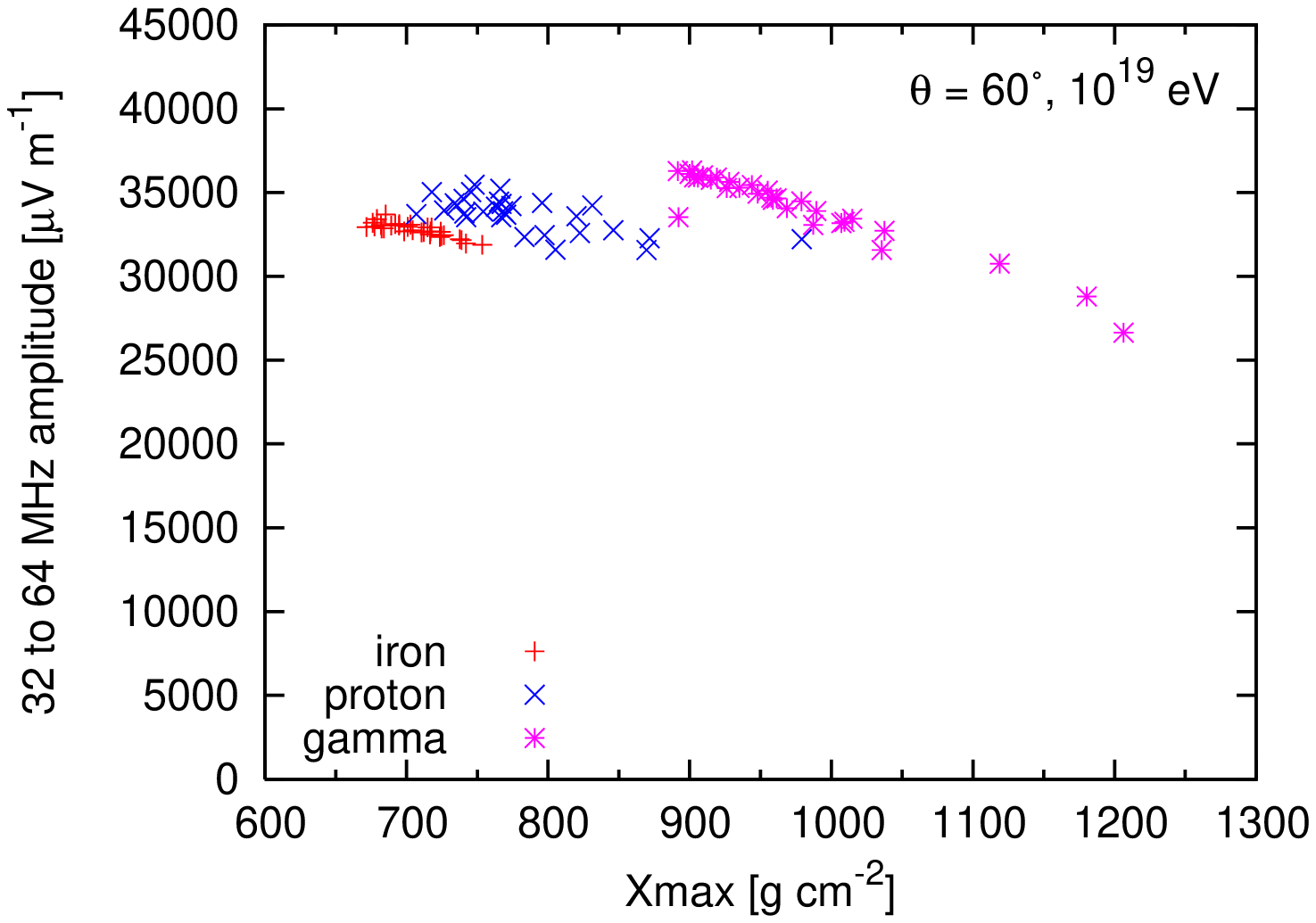}
\caption{\label{flat}Peak filtered radio field strengths of \unit[$10^{19}$]{eV} showers coming from the south as measured in the {\em flat region} \unit[275]{m} north of the shower core.}
\end{minipage} \hspace{1.5pc}
\begin{minipage}{15.5pc}
\includegraphics[angle=270,width=15.5pc]{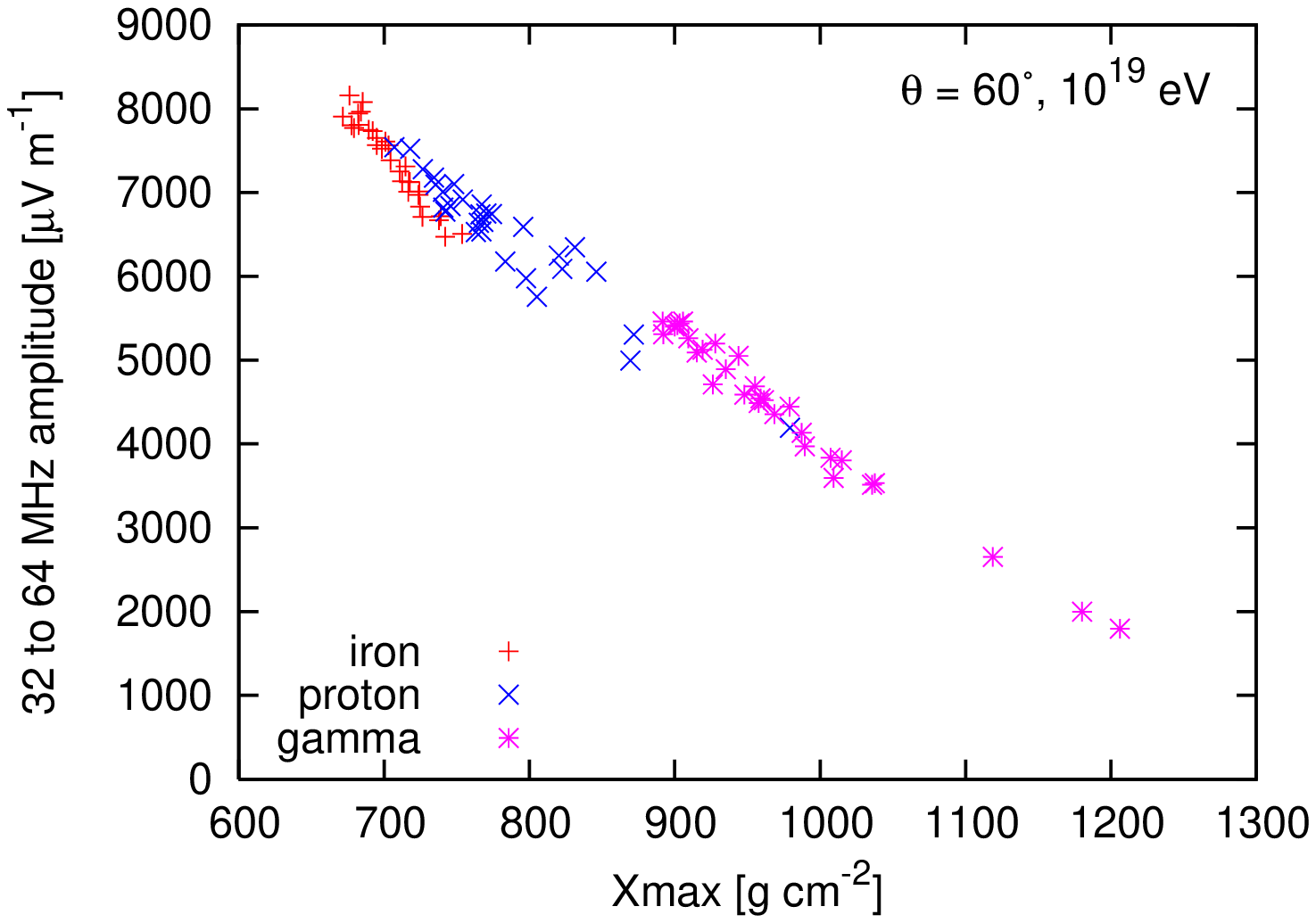}
\caption{\label{steep}Same as Fig.\ \ref{flat}, but in a {\em steep region} \unit[725]{m} north of the shower core.\newline\newline}
\end{minipage}
\end{figure}

A more quantitative view of the \unit[32--64]{MHz} filtered peak amplitudes in this region and their relation to the individual shower $X_{\mathrm{max}}$ values is given in Fig.\ \ref{flat}. The distribution of filtered field strengths is approximately independent of $X_{\mathrm{max}}$ --- and thus almost independent of shower to shower fluctuations. We call the distance region of the intersection, characterised by the minimum in the RMS spread of the amplitudes, {\em flat region} from now on. (The location of this region depends on several parameters, in particular the shower zenith angle and window of observing frequency, but also the chosen normalisation applied to the radio amplitudes, as will be discussed in the following.) Although the amplitude distribution is approximately flat, there are visible ``steps'' between the values for the different types of primary particles. The reason for these ``steps'' is the fact that the radio emission is generated only by the electromagnetic cascade of an air shower, and depending on the type of the primary particle, a different fraction of the energy of the primary particle is transferred to the electromagnetic cascade \citep{SongCaoDawson2000,BarbosaCatalaniChinellato2004}.

At large lateral distances, e.g. around \unit[700]{m}, the lateral profiles for the individual air showers plotted in Fig.\ \ref{lateralslopes} lead to very different filtered electric field strengths. Such a region will be called {\em steep region} from now on. A quantitative look at the filtered electric field strengths as a function of $X_{\mathrm{max}}$ is given in Fig.\ \ref{steep}. In a {\em steep region}, the filtered electric field strength is directly correlated with the $X_{\mathrm{max}}$ of the air shower.

%\begin{figure}[htb]
%\begin{minipage}{15.5pc}
%\centering
%\includegraphics[angle=270,width=20.0pc]{60deg0deg_1e19_rect32to64_obsplot.eps}
%\caption{\label{obsplot}A combined measurement of the \unit[32--64]{MHz} field strengths in a {\em steep region} and a {\em flat region} distinguishes between different primary particle types.}
%\end{minipage} \hspace{1.5pc}
%\begin{minipage}{15.5pc}
%\includegraphics[angle=270,width=15.5pc]{60deg0deg_1e19_lateralslopes_maxamp.eps}
%\caption{\label{bla}Bla.}
%\end{minipage}
%\end{figure}

%If one combines field strengths measured in the {\em flat region} and a {\em steep region} into a scatter plot as depicted in Fig.\ \ref{obsplot}, it becomes clear that the ratio of the radio field strengths in these two different regions contains information on the primary particle type.

\subsection{Energy determination}

With experimental measurements, the arrival direction and core position of an air shower can be deduced from radio-only measurements \citep{ArdouinBelletoileCharrier2006}. The energy of the primary particle, on the other hand, is a priori unknown. The correlations discussed in the previous section can, however, be generalized for a range of energies.

\begin{figure}[htb]
%\begin{minipage}{15.5pc}
\centering
\includegraphics[angle=270,width=20.0pc]{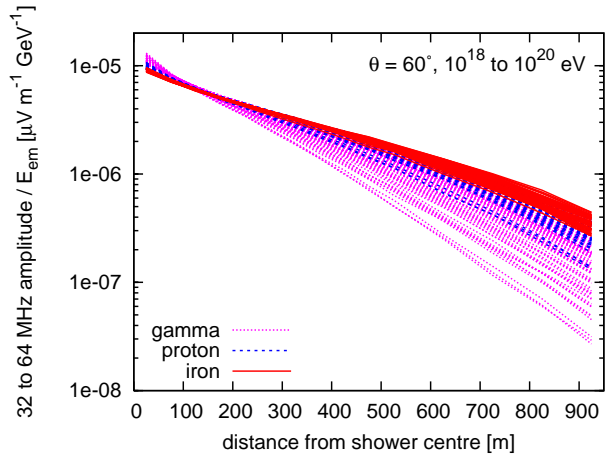}
\caption{\label{lateralslopesallenergies}Lateral distribution of the \unit[32--64]{MHz} filtered peak radio amplitude normalised with the energy deposited in the atmosphere by the electromagnetic cascade. Shown are 180 air showers spanning the energy range from $10^{18}$ to \unit[$10^{20}$]{eV}. The normalised field strength profiles show the same behaviour as those of the monoenergetic showers in Fig.\ \ref{lateralslopes}. In particular, all normalised lateral profiles intersect in a narrow region around $\sim 175\,$m lateral distance.}
%\end{minipage} \hspace{1.5pc}
%\begin{minipage}{15.5pc}
%\includegraphics[angle=270,width=15.5pc]{60deg0deg_1e19_lateralslopes_maxamp.eps}
%\caption{\label{bla}Bla.}
%\end{minipage}
\end{figure}

In Fig.\ \ref{lateralslopesallenergies}, the lateral radio emission profiles of the 180 simulated air showers spanning the energy range from $10^{18}$ to \unit[$10^{20}$]{eV} are shown after filtering to the observing bandwidth of \unit[32--64]{MHz}. To make shower signals comparable we have scaled them with the energy that the electromagnetic cascade of the air shower deposits in the atmosphere. While the absolute electric field amplitudes of the 180 showers span three orders of magnitude, the normalised lateral distribution profiles exhibit the same qualitative behaviour as those of the monoenergetic showers depicted in Fig.\ \ref{lateralslopes}. In particular, there exists again a well-defined region for all showers (and thus all energies), in which the normalised lateral profiles intersect. As will be discussed later, the location of the intersection region and the spread of the normalised field strengths at this distinctive lateral distance depends on the quantity with which the profiles are scaled.

\begin{figure}[htb]
\begin{minipage}{15.5pc}
\includegraphics[angle=270,width=15.5pc]{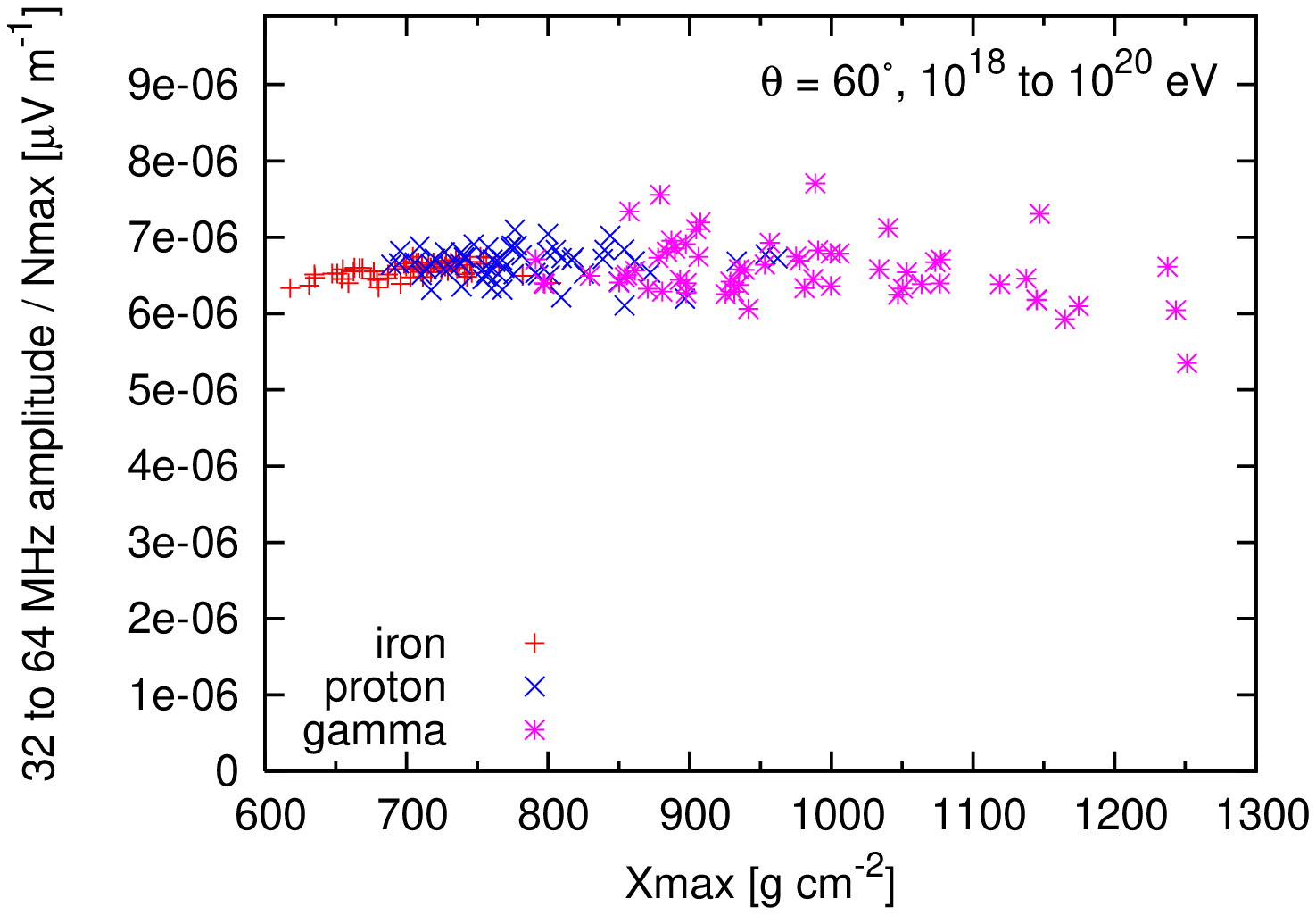}
\caption{\label{flatnormNmax}At \unit[225]{m} north of the shower core, the radio pulse height per $N_{\mathrm{max}}$ over energies of $10^{18}$ to \unit[$10^{20}$]{eV} and all three particle types is approximately constant, i.e., this is a {\em flat region}.}
\end{minipage} \hspace{1.5pc}
\begin{minipage}{15.5pc}
\includegraphics[angle=270,width=15.5pc]{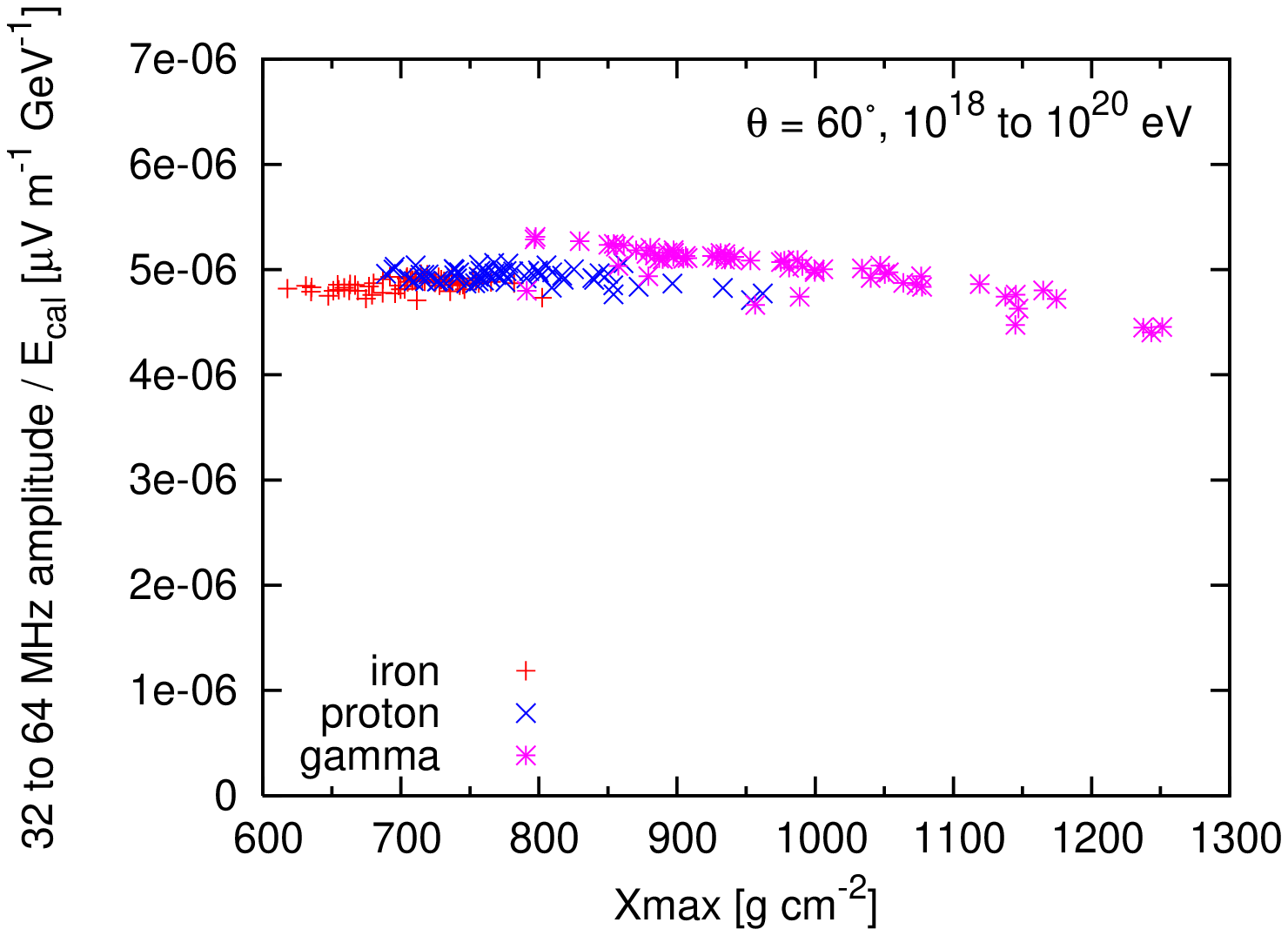}
\caption{\label{flatnormCalorimetric}Normalising the radio signal with the calorimetric energy $E_{\mathrm{cal}}$ of an air shower, yields even better results. The {\em flat region} then lies at \unit[175]{m} north of the shower core.}
\end{minipage}
\end{figure}

Already in \citep{HuegeIcrc2007a}, we have demonstrated that the radio signal correlates with the sum of the number of electrons and positrons in the air shower maximum, $N_{\mathrm{max}}$. This correlation is illustrated in Fig.\ \ref{flatnormNmax}, where the filtered electric field strength per $N_{\mathrm{max}}$ in the appropriate {\em flat region} of all 180 showers with 60$^{\circ}$ zenith angle coming from the south, spanning the energy range from \unit[$10^{18}$]{eV} to \unit[$10^{20}$]{eV}, is plotted against $X_{\mathrm{max}}$. Regardless of the energy and type of the primary particle and the value of $X_{\mathrm{max}}$, the filtered peak electric field strength per $N_{\mathrm{max}}$ yields approximately the same value. In particular, the ``steps'' between the different types of primary particles seen in Fig.\ \ref{flat} have been reduced considerably by the normalisation with the maximum number of particles in the electromagnetic cascade.

Another quantity with which the radio signal could be normalised is the ``calorimetric energy'' in the air shower, $E_{\mathrm{cal}}$. (This is the quantity with which the fluorescence light correlates best, cf.\ \citep{BarbosaCatalaniChinellato2004,RisseHeck2004}.) The calorimetric energy of each simulated air shower is calculated from the total energy deposited in an infinitely thick atmosphere. As demonstrated in Fig.\ \ref{flatnormCalorimetric}, for 60$^{\circ}$ zenith angle showers, the calorimetric energy $E_{\mathrm{cal}}$ has an even better correlation with the radio signal in the appropriate {\em flat region} than $N_{\mathrm{max}}$. The reason is that $E_{\mathrm{cal}}$ fluctuates less from shower to shower than the value of $N_{\mathrm{max}}$.

\begin{figure}[htb]
\begin{minipage}{15.5pc}
\includegraphics[angle=270,width=15.5pc]{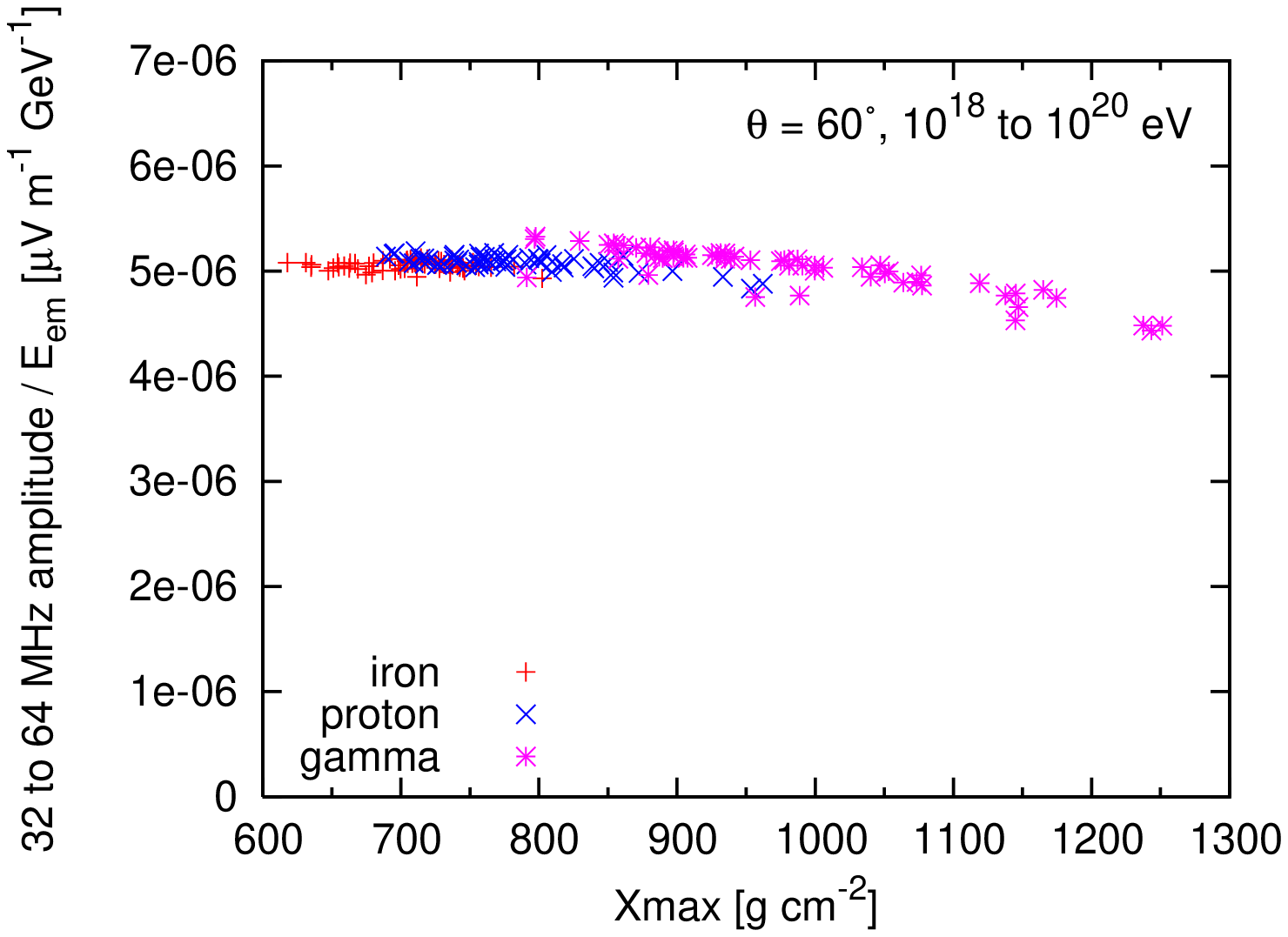}
\caption{\label{flatnorm}For radio pulse heights normalised with $E_{\mathrm{em}}$, the {\em flat region} lies again at a distance of $\sim 175\,$m north of the shower core.\newline}
\end{minipage} \hspace{1.5pc}
\begin{minipage}{15.5pc}
\includegraphics[angle=270,width=15.5pc]{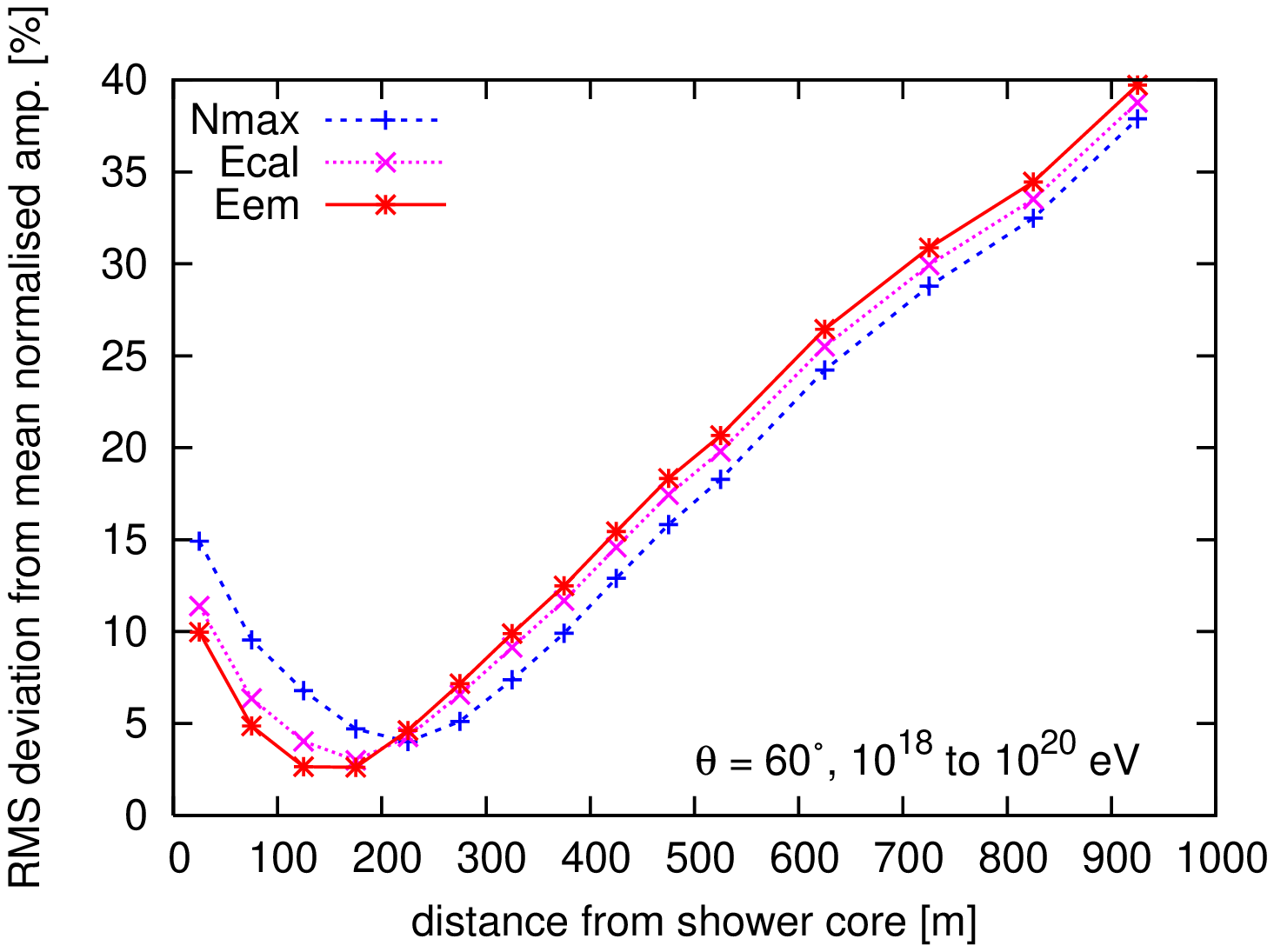}
\caption{\label{normvariants}The lowest deviation from the mean normalised amplitude is reached for the normalisation with $E_{\mathrm{em}}$ with an RMS of $<3$\% in the {\em flat region} at $\sim 175\,$m north.}
\end{minipage}
\end{figure}

From a theoretical point of view, however, the quantity with which the radio signal should correlate best is the ``energy deposited in the atmosphere by the electromagnetic cascade'', $E_{\mathrm{em}}$. In contrast to $E_{\mathrm{cal}}$, $E_{\mathrm{em}}$ neither includes energy deposited by muons or hadrons (which do not contribute to the radio signal) nor takes into account the energy deposited by particles in the ground. The high quality of the correlation with $E_{\mathrm{em}}$ is illustrated in Fig.\ \ref{flatnorm}, which shows the normalised radio amplitudes at \unit[175]{m} north of the shower core, where the corresponding {\em flat region} lies (see also Fig.\ \ref{lateralslopesallenergies}).

In Fig.\ \ref{normvariants} we compare the width of the correlation with the three discussed quantities, $N_{\mathrm{max}}$, $E_{\mathrm{cal}}$ and $E_{\mathrm{em}}$. As mentioned earlier, the location of the {\em flat region}, i.e., the minimum of the RMS distribution, depends on the chosen normalisation. In addition, it becomes clear that $E_{\mathrm{em}}$ yields the best correlation, with an RMS deviation from the mean normalised amplitude of $<$3\% at a ground distance of $\sim 175\,$m. We therefore apply the normalisation of the radio signal with $E_{\mathrm{em}}$ in the following.

\begin{figure}[htb]
\begin{minipage}{15.5pc}
\includegraphics[angle=270,width=15.5pc]{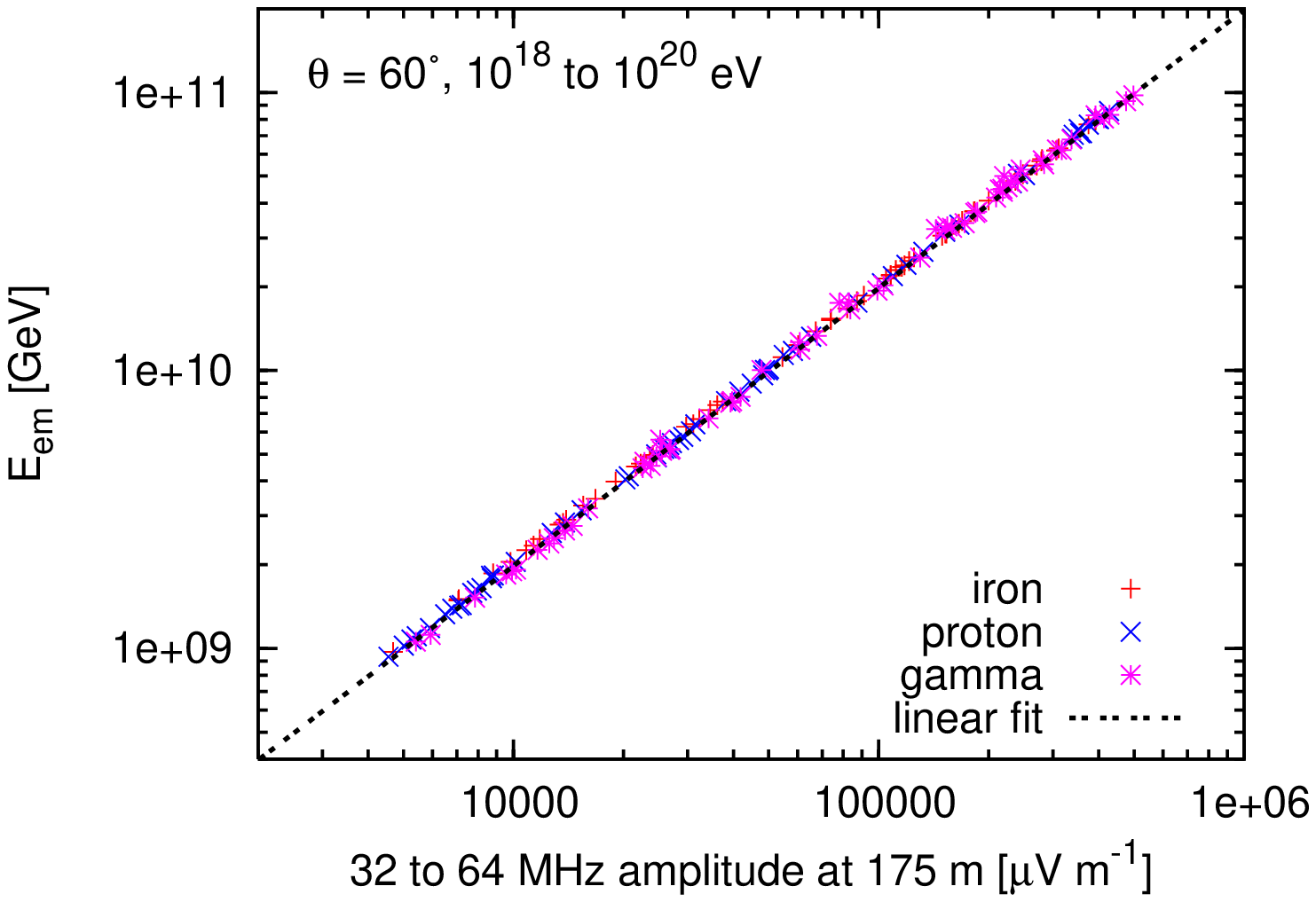}
\caption{\label{energy}A measurement in the {\em flat region} directly yields the shower $E_{\mathrm{em}}$ on a shower-to-shower basis via a linear relation. The fit paramater is $a=197900$.}
\end{minipage} \hspace{1.5pc}
\begin{minipage}{15.5pc}
\includegraphics[angle=270,width=15.5pc]{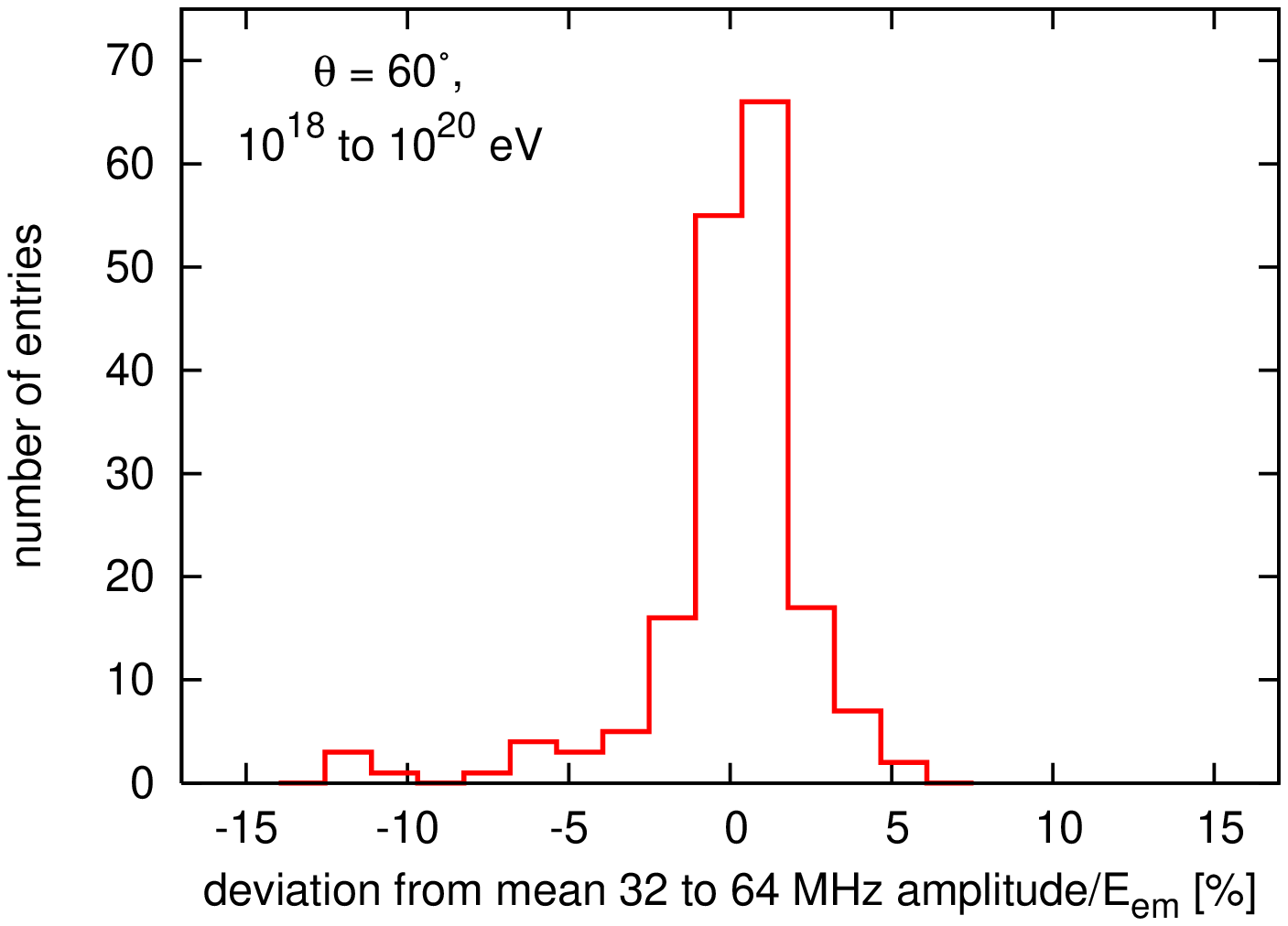}
\caption{\label{flatrms}The RMS spread in peak field strengths per $E_{\mathrm{em}}$ in the {\em flat region} over all energies and particle types amounts to $<$3\%.}
\end{minipage}
\end{figure}

Once the value of the normalised, energy-independent radio amplitude per $E_{\mathrm{em}}$ in the {\em flat region} is known for a given geometry and observing frequency --- either from simulations or from measurements --- a measurement of the peak filtered radio amplitude $A_{\mathrm{flat}}$ in the {\em flat region} can be used to immediately read off the value of $E_{\mathrm{em}}$ in a linear relation,
\begin{equation}
E_{\mathrm{em}}=a\ A_{\mathrm{flat}}\ \mu\mathrm{V}^{-1}\ \mathrm{m}\ \mathrm{GeV},
\end{equation}
where $a$ is a shower geometry dependent parameter, see Fig.\ \ref{energy}. In Fig.\ \ref{flatrms}, the distribution of the \unit[32--64]{MHz} filtered peak radio amplitudes per $E_{\mathrm{em}}$ around their mean value is illustrated. As discussed earlier, the RMS of this distribution amounts to $<$3\%. In principle, i.e. not considering experimental uncertainties, the $E_{\mathrm{em}}$ value can therefore be inferred with a theoretical uncertainty of $<$3\%. It should be stressed that this theoretical accuracy would be achievable on a shower-to-shower basis, unhindered by shower-to-shower fluctuations and without a priori knowledge of the mass of the primary particle. The derived value of $E_{\mathrm{em}}$ can in turn be related to the energy of the primary particle with an additional uncertainty of the order $\sim$5-7\% \citep{AlvarezMunizEngelGaisser2004}.

\subsection{Composition determination}

\begin{figure}[htb]
\begin{minipage}{15.5pc}
\includegraphics[angle=270,width=15.5pc]{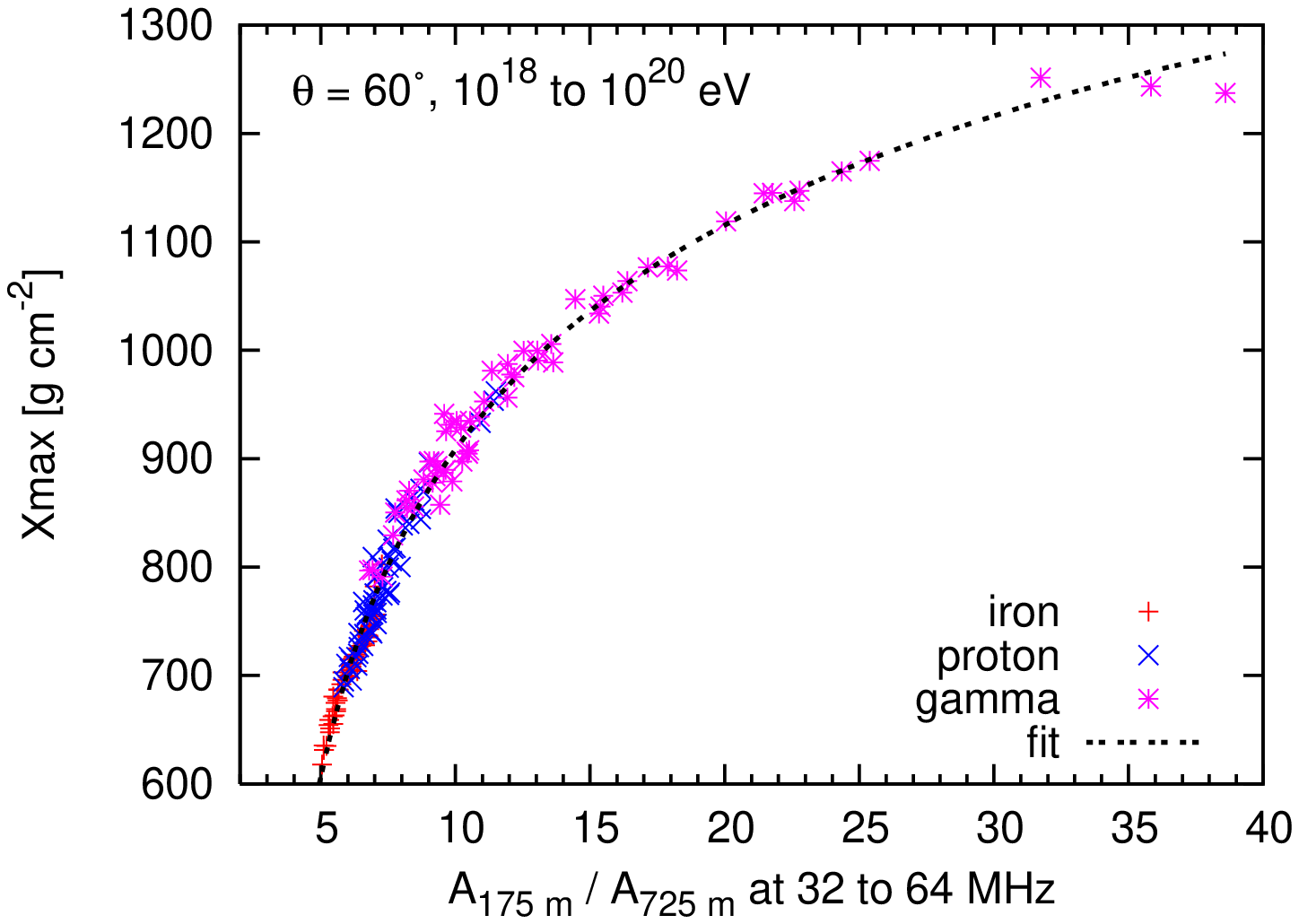}
\caption{\label{composition}The ratio of peak field strengths in the {\em flat region} and {\em steep region} yields direct information on the $X_{\mathrm{max}}$ value of an individual air shower. The fit parameters are $a=856.1$, $b=0.3149$ and $c=0.4340$.}
\end{minipage} \hspace{1.5pc}
\begin{minipage}{15.5pc}
\includegraphics[angle=270,width=15.5pc]{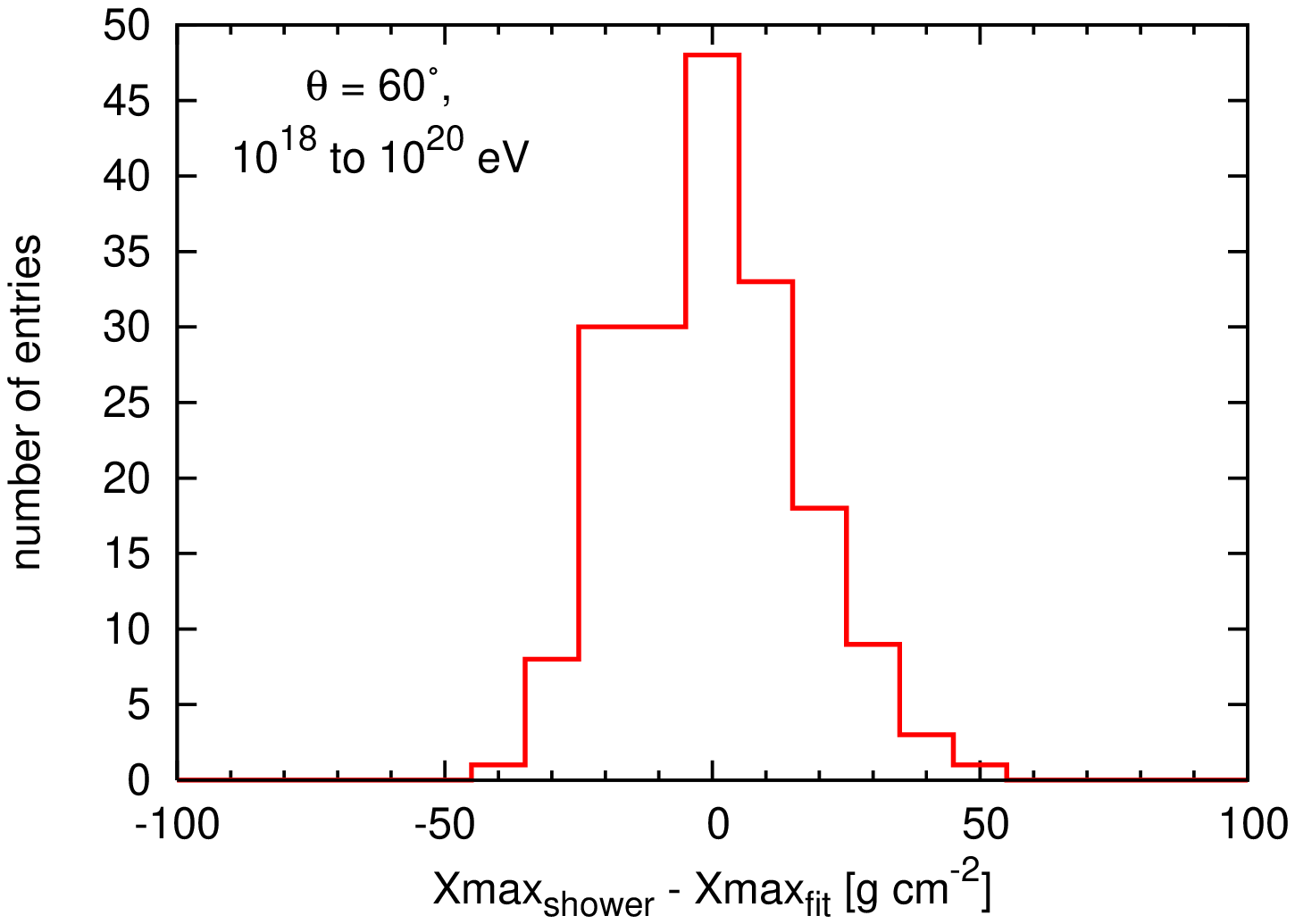}
\caption{\label{compositionrms}The RMS spread of the individual shower $X_{\mathrm{max}}$ values around the values calculated with the fit function amounts to 15.9$\,$g$\,$cm$^{-2}$.\newline\newline}
\end{minipage}
\end{figure}

%In analogy to the scatter plot shown in Fig.\ \ref{obsplot}
The ratio of the measured field strengths in the {\em flat region} and {\em steep region} can be used to infer the $X_{\mathrm{max}}$ value on a shower-to-shower basis, which can in turn be related to the mass of the primary particle. This is demonstrated in Fig.\ \ref{composition}. The correlation of $X_{\mathrm{max}}$ with the ratio $A_{\mathrm{flat}}/A_{\mathrm{steep}}$ can be described well with the fit function
\begin{equation}
X_{\mathrm{max}} = a \left[\ln\left(b\ \frac{A_{\mathrm{flat}}}{A_{\mathrm{steep}}}\right)\right]^{c}~ \mathrm{g\ cm}^{-2}.
\end{equation}
The deviations of the $X_{\mathrm{max}}$ values of individual showers from this fit are depicted in Fig.\ \ref{compositionrms}. The RMS spread of these deviations for the 180 showers shown here amounts to 15.9$\,$g$\,$cm$^{-2}$.

Note that this spread only accounts for intrinsic shower-to-shower fluctuations and does not include any measurement errors. For comparison, fluorescence telescopes, which measure $X_{\mathrm{max}}$ directly, reach an experimental resolution of the order of 35 to \unit[20]{g cm$^{-2}$} \citep{AbbasiAbuZayyadArchbold2005,DawsonICRC2007}.

\subsection{Influence of observing frequency}

\begin{figure}[htb]
\begin{minipage}{15.5pc}
\includegraphics[angle=270,width=15.5pc]{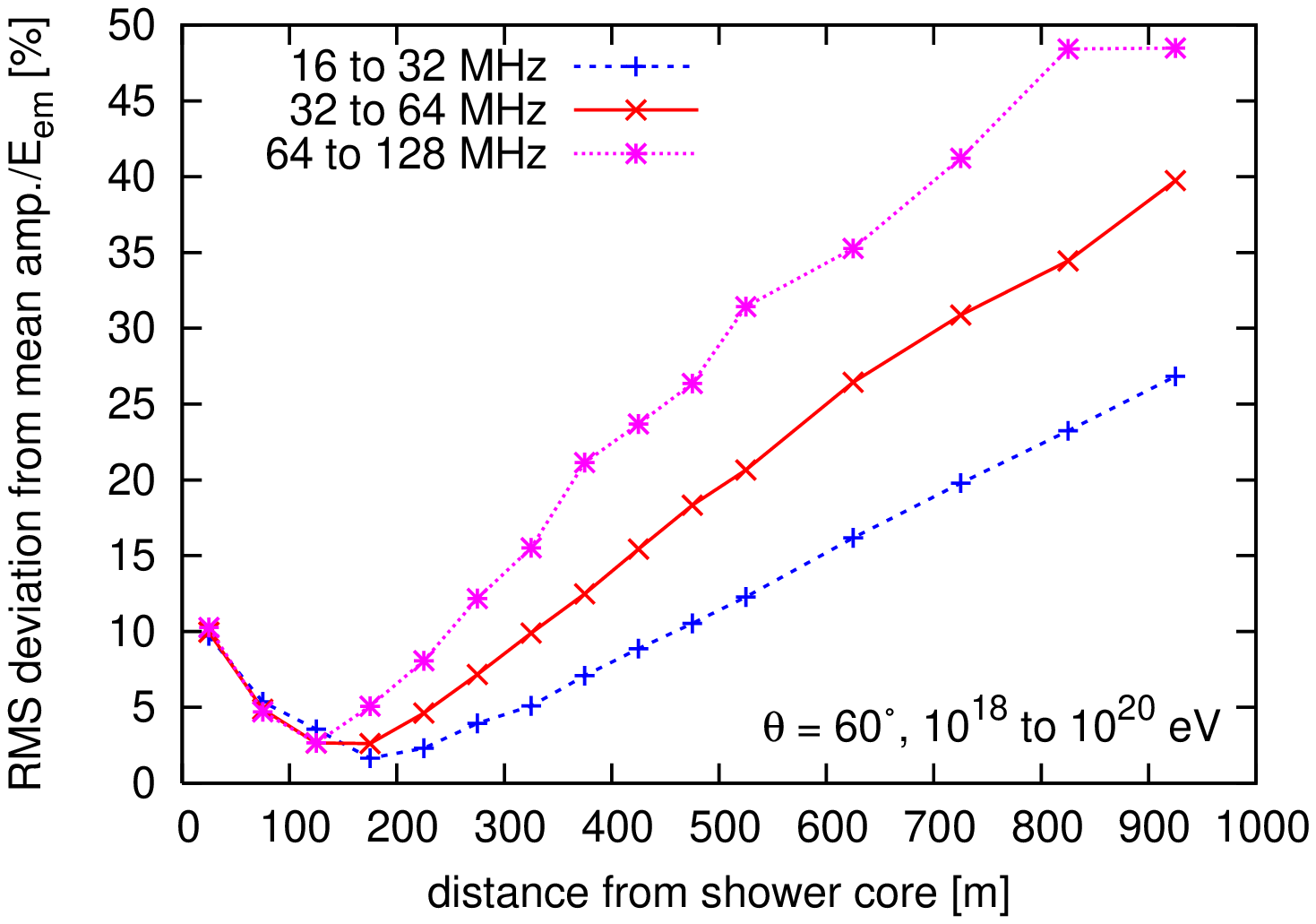}
\caption{\label{rmsopt}RMS spread of the peak field strengths per $E_{\mathrm{em}}$ as a function of ground distance from the shower core in the azimuthal direction defined by the shower axis for observers at \unit[1400]{m} above sea level.}
\end{minipage} \hspace{1.5pc}
\begin{minipage}{15.5pc}
\includegraphics[angle=270,width=15.5pc]{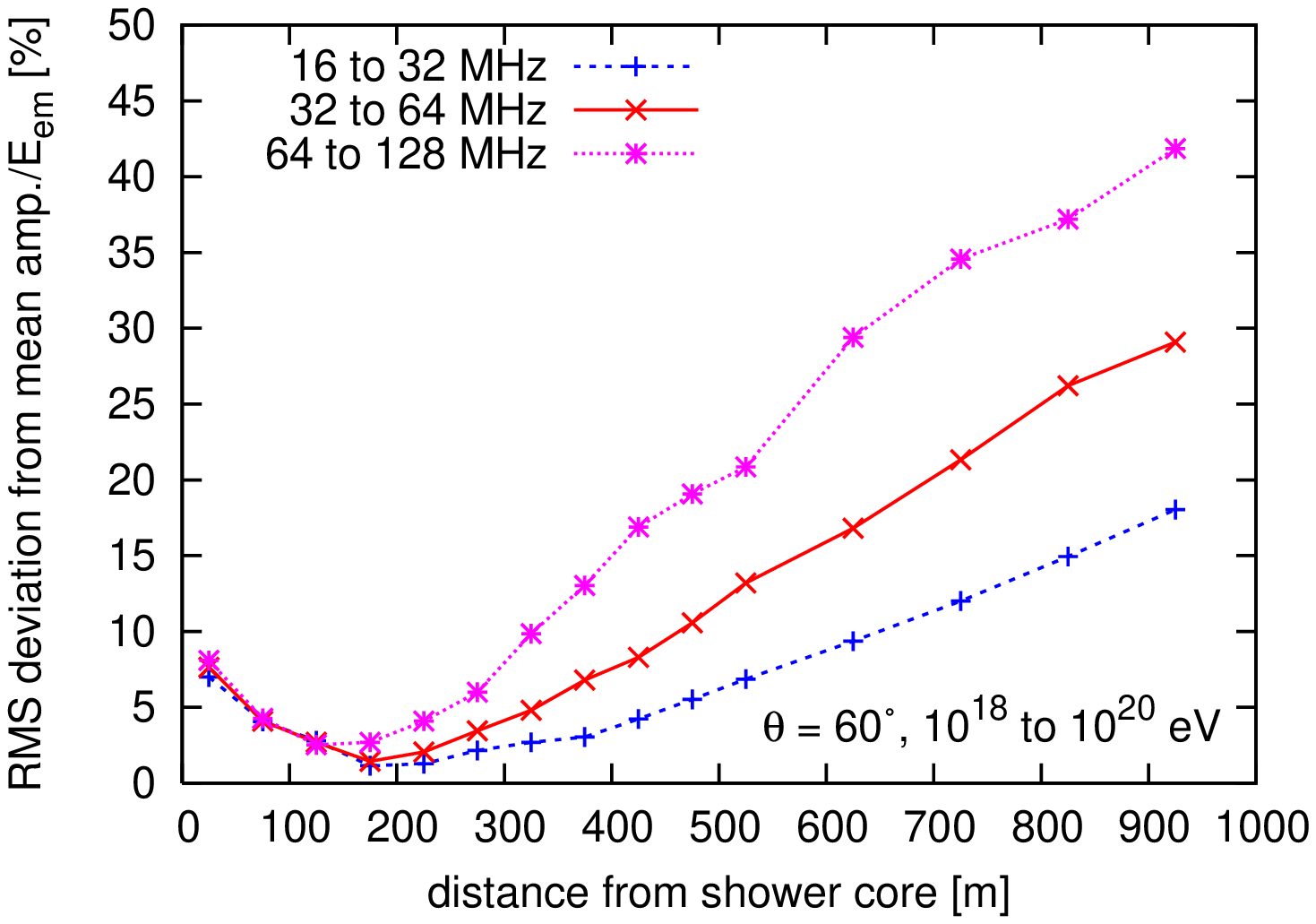}
\caption{\label{sealevel60deg}Same as Fig.\ \ref{rmsopt} but for observers at sea level. The {\em flat regions} shift to slightly larger distances.\newline\newline\newline}
\end{minipage}
\end{figure}

The characteristic lateral distance at which the {\em flat region} is located depends strongly on the zenith angle of the air shower. The frequency window of the radio observations, however, also influences its location. For the case of $60^{\circ}$ zenith angle air showers and normalisation with $E_{\mathrm{em}}$, this is illustrated in Fig.\ \ref{rmsopt}. A clear minimum of the RMS with values $<$3\% can be deduced for each frequency band separately. The corresponding distance of the {\em flat region} amounts to the aforementioned $\sim 175\,$m in case of the \unit[32--64]{MHz} band. For higher frequencies, it lies at smaller distances, while for lower frequencies, it shifts to larger distances. From an experimental point of view, larger distances are favourable because they require less densely equipped radio antenna arrays. Multi-frequency observations including lower observing frequencies would thus allow an even better exploitation of measurements in the {\em flat} and {\em steep regions} for the determination of $E_{\mathrm{em}}$ and $X_{\mathrm{max}}$.

\subsection{Influence of observation altitude}

The position of the {\em flat region} is also influenced by the observer's height above sea level. This is due to the fact that the geometrical distance from the shower maximum to the radio antennas changes with the observer position, which influences the lateral distributions of the radio emission \citep{HuegeFalcke2005b}.

In Fig.\ \ref{sealevel60deg}, the RMS deviation from the mean filtered radio amplitude per $E_{\mathrm{em}}$ is shown for observers at sea level, in contrast to the \unit[1400]{m} above sea level used as standard value in this analysis. While the difference to an observer at \unit[1400]{m} above sea level (Fig.\ \ref{rmsopt}) is not dramatic for 60$^{\circ}$ zenith angle showers, there is a clear trend for the flat regions to move to larger lateral distances, as can be expected from the flattening of the lateral profiles.

\subsection{Influence of observer azimuth angle}

\begin{figure}[htb]
\centering
\includegraphics[angle=270,width=15.5pc]{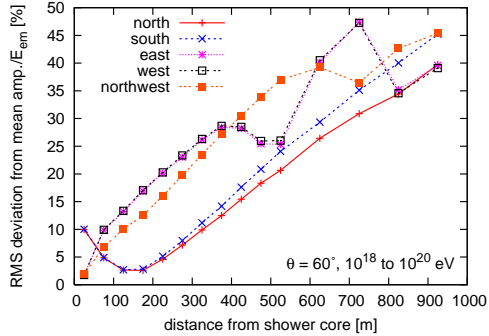}
\caption{\label{rmsoptdirectional}RMS spread of the \unit[32--64]{MHz} peak field strengths per $E_{\mathrm{em}}$ for different observer directions relative to the shower core. The local minima are dependent on the window of observing frequency and thus related to the bandpass filtering.}
\end{figure}

Radio emission from inclined air showers exhibits a significant asymmetry for observer positions along the azimuthal direction given by the air shower axis and positions perpendicular to the air shower axis. For inclined air showers, part of this asymmetry is caused by geometrical effects and can be taken out by changing from ground coordinates to shower coordinates \citep{HuegeFalcke2005b}. The geosynchrotron model, however, also predicts an intrinsic asymmetry that is present even for vertical air showers \citep{HuegeUlrichEngel2007a}. In Fig.\ \ref{rmsoptdirectional}, we illustrate the RMS spread of the \unit[32--64]{MHz} filtered peak amplitudes per $E_{\mathrm{em}}$ for observers along different azimuthal directions in case of air showers coming from the south. The behaviour is almost identical for observers along the two azimuthal directions defined by the shower axis, in this case to the north and to the south, except for a small asymmetry caused by the inclination of the geomagnetic field.

For an observer in the north-west, i.e.\ at 45$^{\circ}$ azimuth to the shower axis, the {\em flat region} shifts to a very small lateral distance, which is no longer resolved in the simulations. This is the behaviour expected from the model-predicted asymmetries of the radio ``footprint''. In addition, there is a local minimum in the RMS distribution around $\sim 725\,$m. On a closer look, such additional local minima are also visible in the north direction, especially at \unit[64--128]{MHz} (cf.\ Fig.\ \ref{rmsopt}). These local minima are dependent on the window of observing frequency and therefore related to interference effects arising from the bandpass filtering.

For the east or west directions, i.e.\ perpendicular to the shower axis, the {\em flat region} also lies at distances too small to be resolved by the simulations. At the same time, additional prominent local minima occur at large distances.

From an experimental point of view, large lateral distances for the {\em flat region} are preferred, and measurements would thus be easiest in the azimuthal directions given by the air shower axis.

\subsection{Influence of shower azimuth angle}

\begin{figure}[htb]
\begin{minipage}{15.5pc}
\includegraphics[angle=270,width=15.5pc]{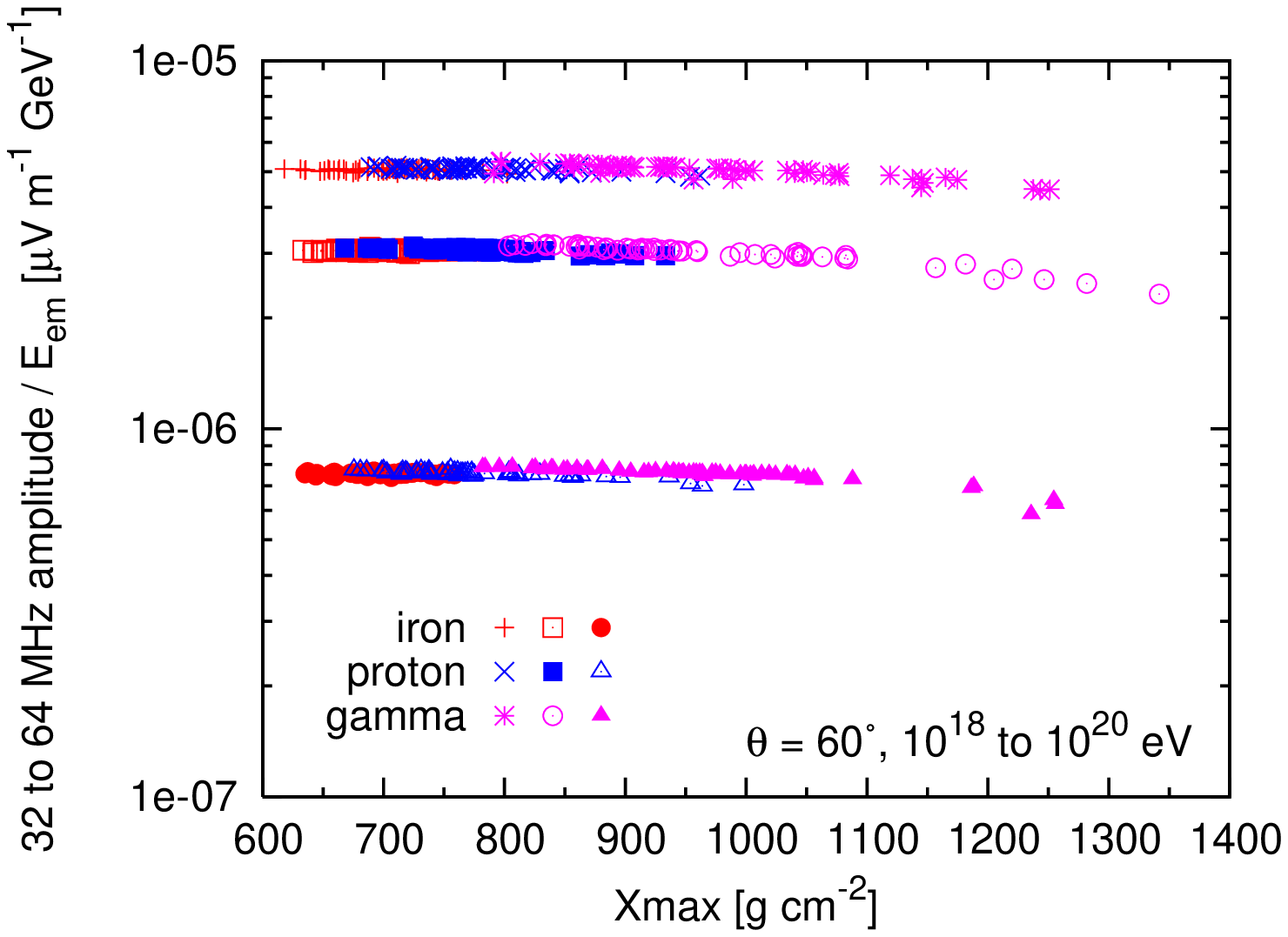}
\caption{\label{flatnormall}For each of the shower azimuth angles, the peak filtered electric field strength per $E_{\mathrm{em}}$ is constant over all energies and particle types at a distance of \unit[175]{m} from the shower core.}
\end{minipage} \hspace{1.5pc}
\begin{minipage}{15.5pc}
\includegraphics[angle=270,width=15.5pc]{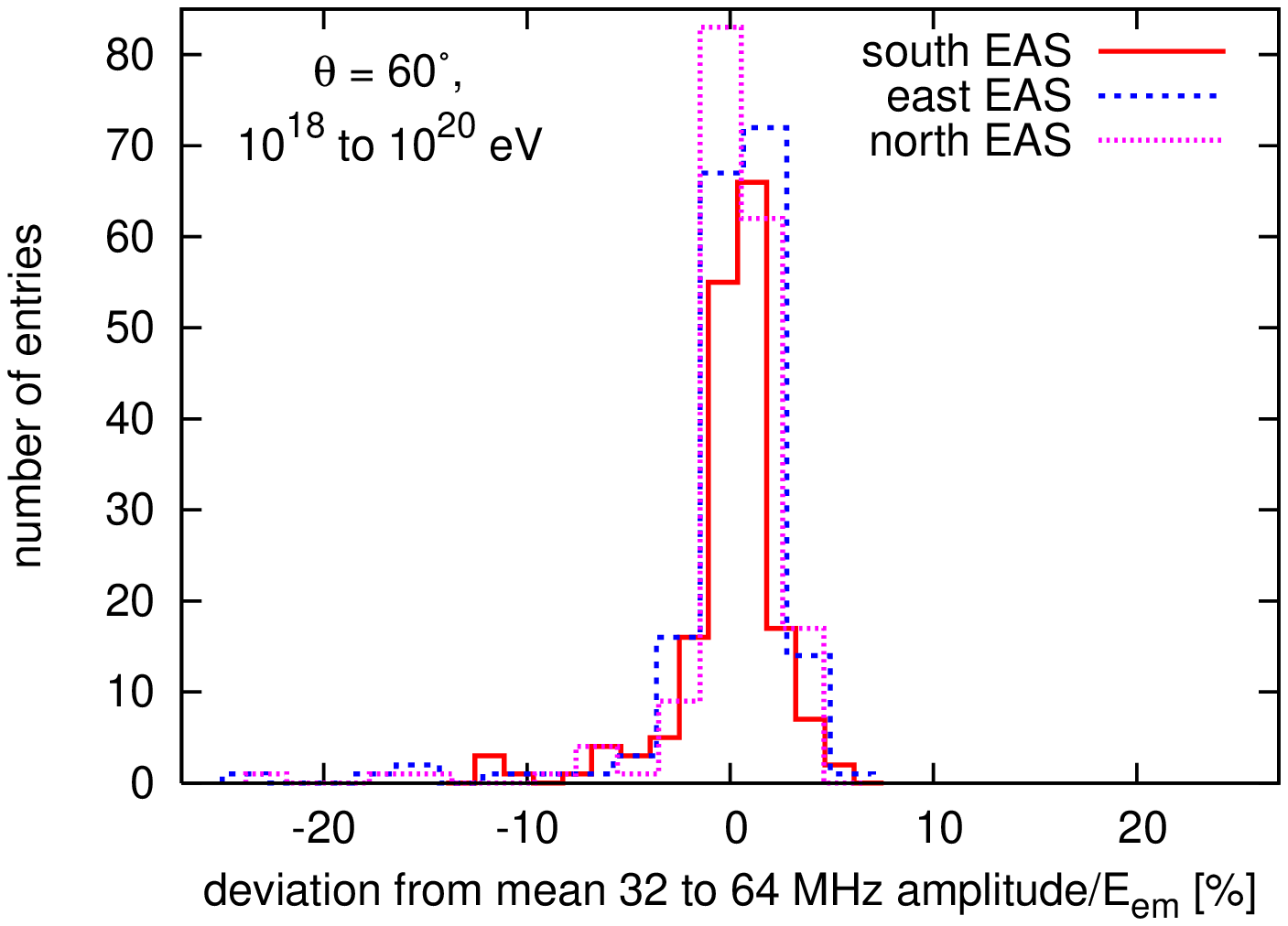}
\caption{\label{flatrmsall}Regardless of shower azimuth, the spread of the peak filtered electric field strength per $E_{\mathrm{em}}$ in the {\em flat region} corresponds to only $\sim 3$\%.\newline}
\end{minipage}
\end{figure}

So far, we have only treated the case of 60$^{\circ}$ zenith angle showers coming from the south. The qualitative behaviour of the radio signal, however, is independent of the shower azimuth angle. This is illustrated in Fig.\ \ref{flatnormall}, where the peak filtered radio amplitude per $E_{\mathrm{em}}$ in the {\em flat region} is plotted for the three sets of air showers coming from the south, east and north. (The {\em flat region} is then, correspondingly, found at $\sim 175\,$m to the north, west and south of the shower core.) While the absolute value of the filtered peak radio amplitude per $E_{\mathrm{em}}$ varies with the angle between the shower axis and the Earth's magnetic field (called {\em geomagnetic angle}, corresponding to 67.3$^{\circ}$, 107.6$^{\circ}$ and 172.7$^{\circ}$ for the three cases), the values for a given geometry are again approximately equal over all showers. Once this constant value is known for a given geometry --- from simulations or measurements --- the $E_{\mathrm{em}}$ determination can again be performed with a measurement in the {\em flat region}. As illustrated in Fig.\ \ref{flatrmsall}, the RMS spread is still $<$3\% in all three cases.

\begin{figure}[htb]
\centering
\includegraphics[angle=270,width=20.0pc]{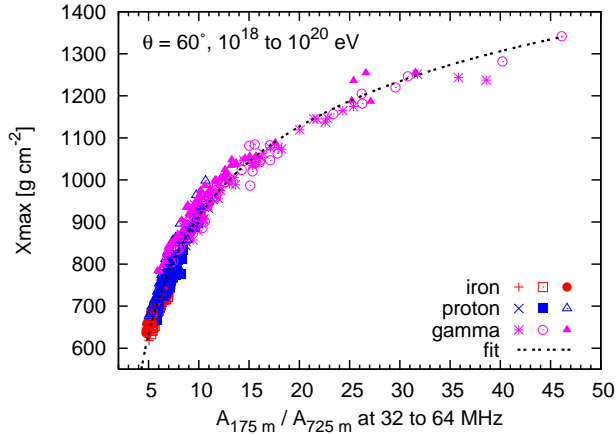}
\caption{\label{compositionall}The correlation of $X_{\mathrm{max}}$ with the ratio of filtered peak amplitudes in the {\em flat region} and {\em steep region} is (to first order) independent of the shower azimuth angle. The fit parameters are $a=789.2$, $b=0.3813$ and $c=0.5025$. The RMS deviation of the individual $X_{\mathrm{max}}$ values from the fit amounts to 25.8$\,$g$\,$cm$^{-2}$.}
\end{figure}

The determination of $X_{\mathrm{max}}$ values via the ratio of filtered radio amplitudes in the {\em flat region} and {\em steep region} also works regardless of the shower azimuth angle. As the calculation of this ratio removes the absolute strength of the radio signals, the correlation of $X_{\mathrm{max}}$ with this ratio is (to first order) universal and independent of the shower azimuth angle. This is illustrated in Fig.\ \ref{compositionall}, which contains all 540~showers of the three geometries in the same diagram. (The change of the fit parameters and RMS compared to Fig.\ \ref{composition}, however, illustrates that there are minor systematic differences between the three geometries.)

\section{The 45$^{\circ}$ zenith angle case}

\begin{figure}[htb]
\begin{minipage}{15.5pc}
\centering
\includegraphics[angle=270,width=15.5pc]{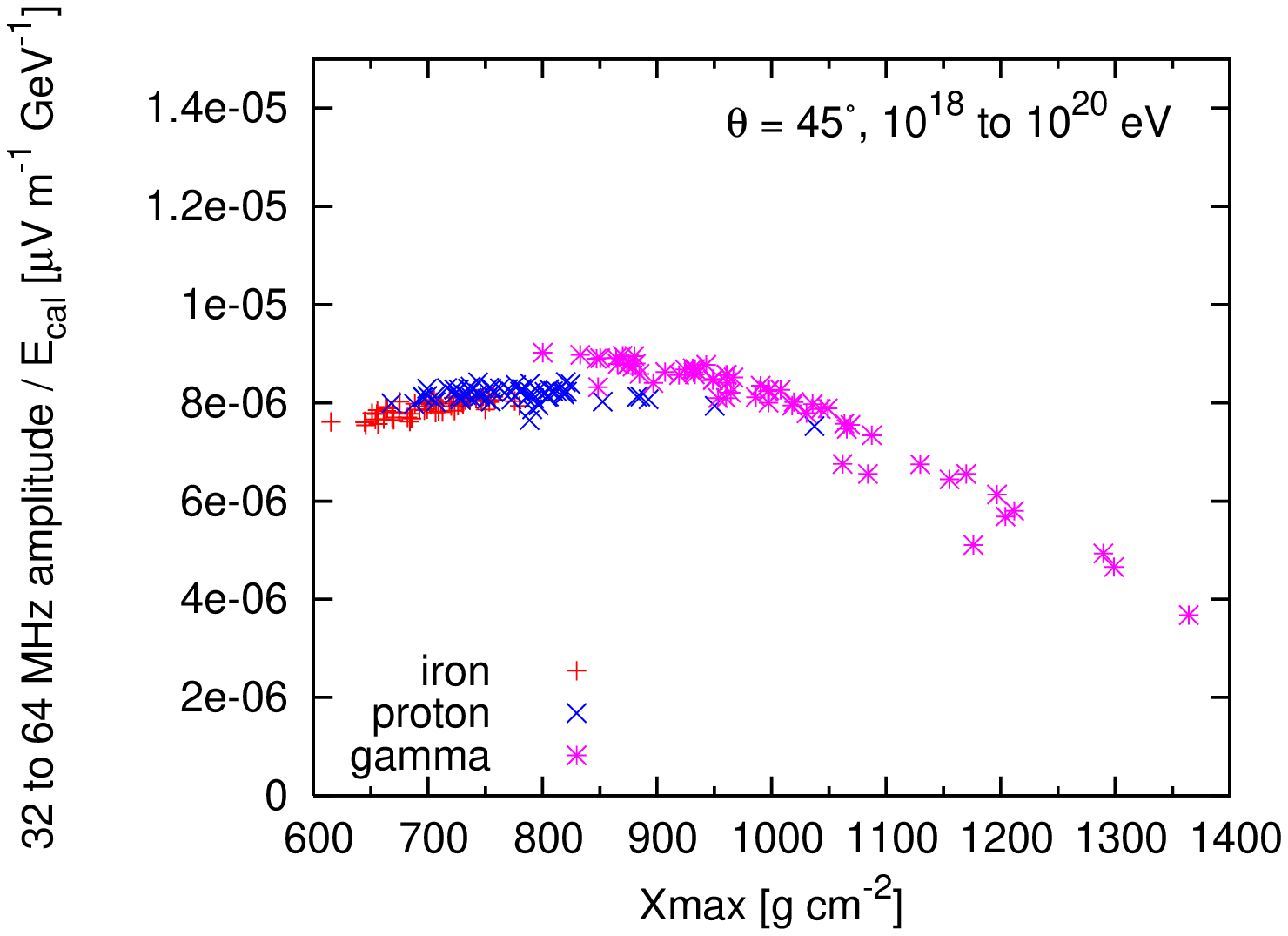}
\caption{\label{45degnormcalorimetric}For 45$^{\circ}$ zenith angle, the normalisation of filtered peak radio amplitudes with $E_{\mathrm{cal}}$ does not work well for deeply penetrating gamma-ray induced showers caused by the LPM effect.}
\end{minipage} \hspace{1.5pc}
\begin{minipage}{15.5pc}
\centering
\includegraphics[angle=270,width=15.5pc]{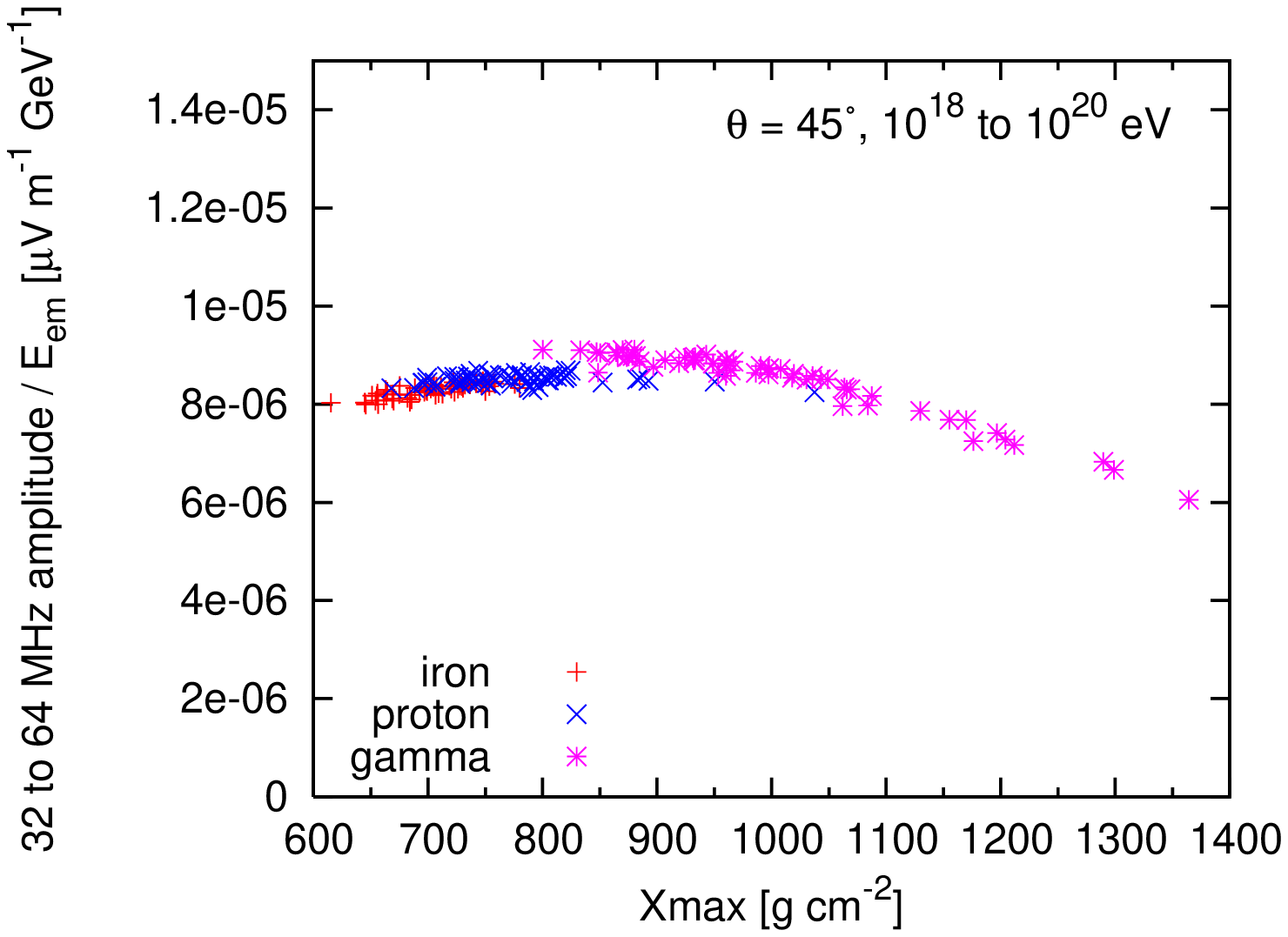}
\caption{\label{45degnormemcal}Normalisation using $E_{\mathrm{em}}$ works better than with $E_{\mathrm{cal}}$, as the energy deposit in the ground is excluded.\newline\newline}
\end{minipage}
\end{figure}

When going from 60$^{\circ}$ zenith angle to 45$^{\circ}$ zenith angle, the qualitative behaviour of the radio signal stays the same, but the lateral distance scales get significantly smaller. One effect which becomes, however, more important, is the inclusion or exclusion of energy deposited in the ground. For 45$^{\circ}$ zenith angle, the ground at \unit[1400]{m} above sea level corresponds to a slant atmospheric depth of $\sim 1240\,$g$\,$cm$^{-2}$. Especially for photon-induced showers that penetrate very deep due to the LPM effect and have $X_{\mathrm{max}}$ values of $>1200\,$g$\,$cm$^{-2}$, a large fraction of the energy is then deposited in the ground. The importance of this aspect is illustrated by the comparison of Fig.\ \ref{45degnormcalorimetric} and Fig.\ \ref{45degnormemcal}. Signals from showers with very high values of $X_{\mathrm{max}}$ are weighted down too much if $E_{\mathrm{cal}}$ is used for normalisation, i.e., the energy deposit in the ground is included. The normalisation with $E_{\mathrm{em}}$, on the other hand, works better, as the energy deposit in the ground is excluded.

\begin{figure}[htb]
\centering
\includegraphics[angle=270,width=15.5pc]{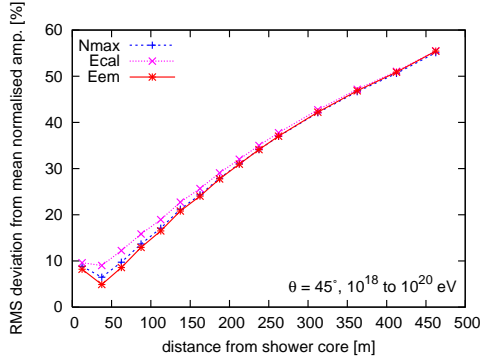}
\caption{\label{rmsopt45degallnorms}RMS spread of the \unit[32--64]{MHz} filtered radio peak amplitudes normalised to $N_{\mathrm{max}}$, $E_{\mathrm{cal}}$ and $E_{\mathrm{em}}$, respectively, as a function of ground distance from the shower core in the azimuthal direction given by the shower axis for 45$^{\circ}$ zenith angle.}
\end{figure}

The quality of the different normalisations (with $N_{\mathrm{max}}$, $E_{\mathrm{cal}}$ and $E_{\mathrm{em}}$) is further illustrated in Fig.\ \ref{rmsopt45degallnorms}. In contrast to the 60$^{\circ}$ zenith angle case, the normalisation with $E_{\mathrm{cal}}$ is now worse than that with $N_{\mathrm{max}}$.

\begin{figure}[htb]
\begin{minipage}{15.5pc}
\centering
\includegraphics[angle=270,width=15.5pc]{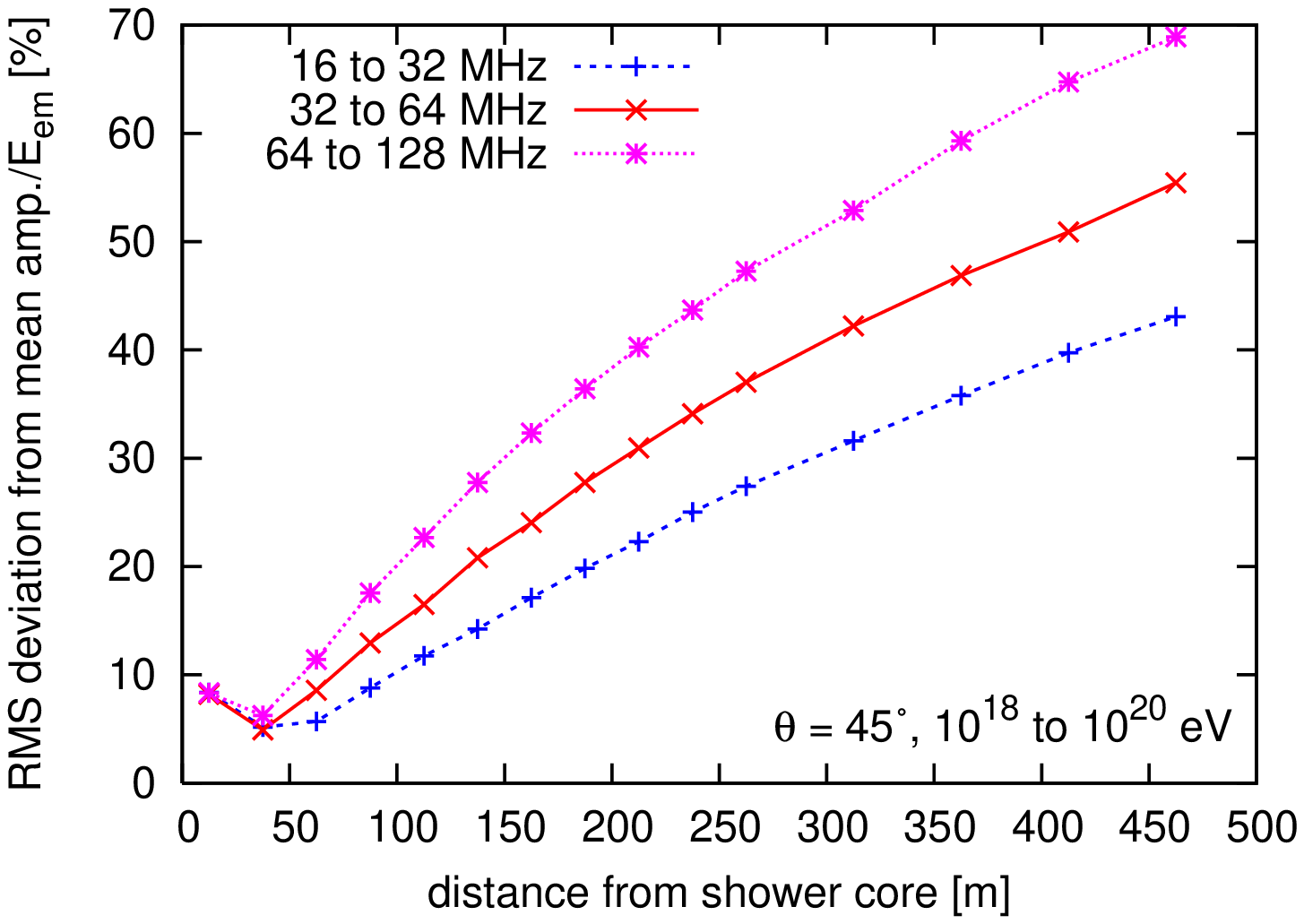}
\caption{\label{rmsopt45deg}RMS spread of the filtered radio peak amplitude per $E_{\mathrm{em}}$ as a function of ground distance from the shower core in the azimuthal direction given by the shower axis for 45$^{\circ}$ zenith angle showers and observers at \unit[1400]{m} above sea level.}
\end{minipage} \hspace{1.5pc}
\begin{minipage}{15.5pc}
\centering
\includegraphics[angle=270,width=15.5pc]{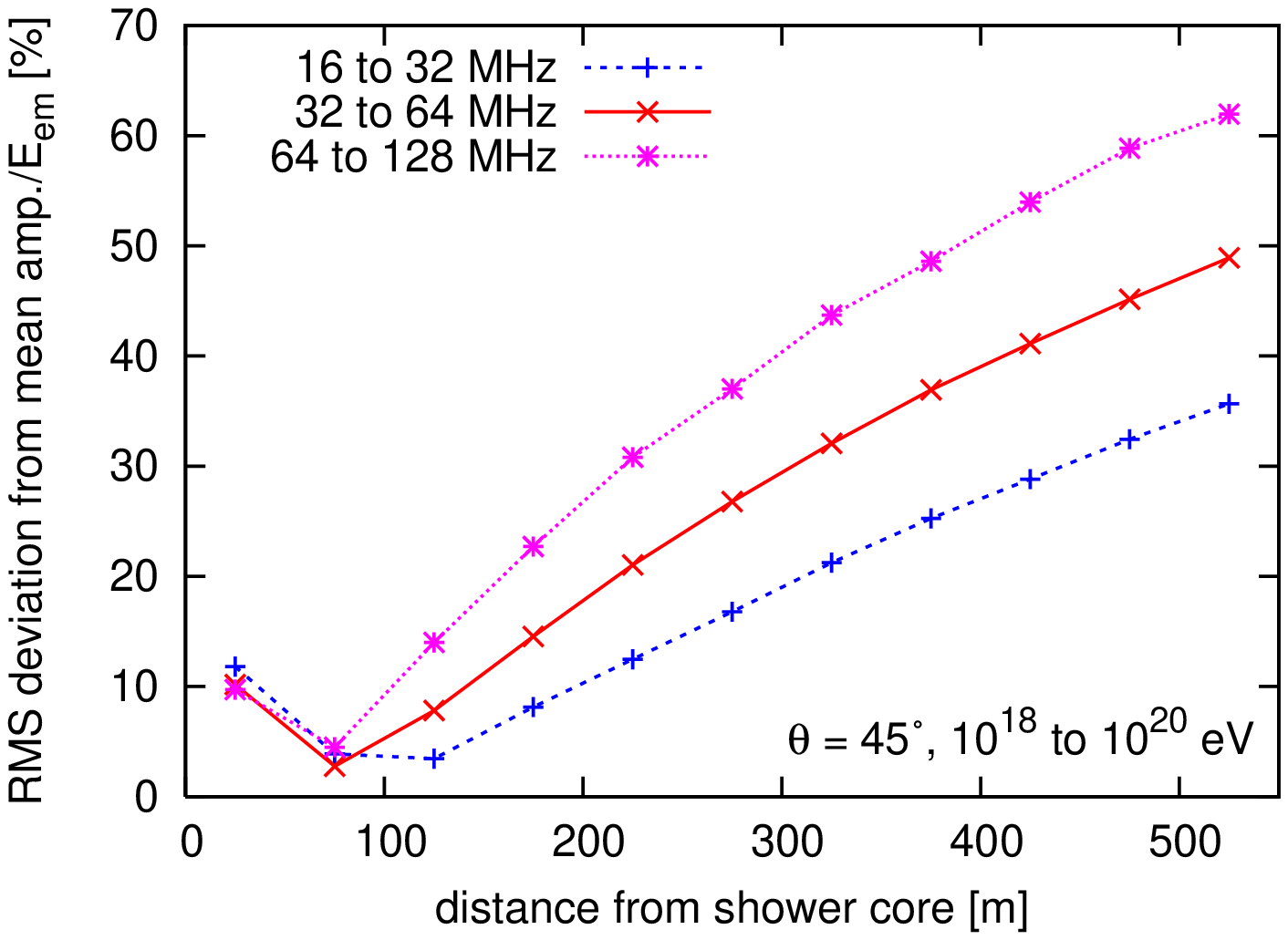}
\caption{\label{sealevel45deg}Same as Fig.\ \ref{rmsopt45deg} but for observers at sea level.\newline\newline\newline\newline\newline}
\end{minipage}
\end{figure}

If the normalisation with $E_{\mathrm{em}}$ is kept, the qualitative behaviour is the same as in the 60$^{\circ}$ zenith angle case. This is illustrated by Fig. \ref{rmsopt45deg}, which again shows the RMS spread of the peak filtered radio amplitude per $E_{\mathrm{em}}$ as a function of lateral distance along the north (for showers coming from the south) for different windows of observing frequency. A clear minimum in RMS, caused by the intersection of the normalised lateral radio profiles, is again present, but the corresponding {\em flat region} now lies at distances around \unit[50]{m}, depending on the observing frequency. The achievable $E_{\mathrm{em}}$ resolution is $\sim 5$\%.

When comparing the radio emission for observers at \unit[1400]{m} above sea level (Fig.\ \ref{rmsopt45deg}) and at sea level (Fig.\ \ref{sealevel45deg}), it becomes obvious that the changes in the 45$^{\circ}$ zenith angle case are much more dramatic than for 60$^{\circ}$ zenith angle, as expected from the behaviour of the lateral profiles with the geometric distance between observer and shower maximum \citep{HuegeFalcke2005b}. For observers at sea level, the {\em flat region} lies at ground distances between 50 and \unit[150]{m}, which comprises a suitable range for densely equipped antenna arrays such as LOPES \citep{HornefferArena2005} or LOFAR \citep{FalckevanHaarlemdeBruyn2007}.

\begin{figure}[htb]
\begin{minipage}{15.5pc}
\includegraphics[angle=270,width=15.5pc]{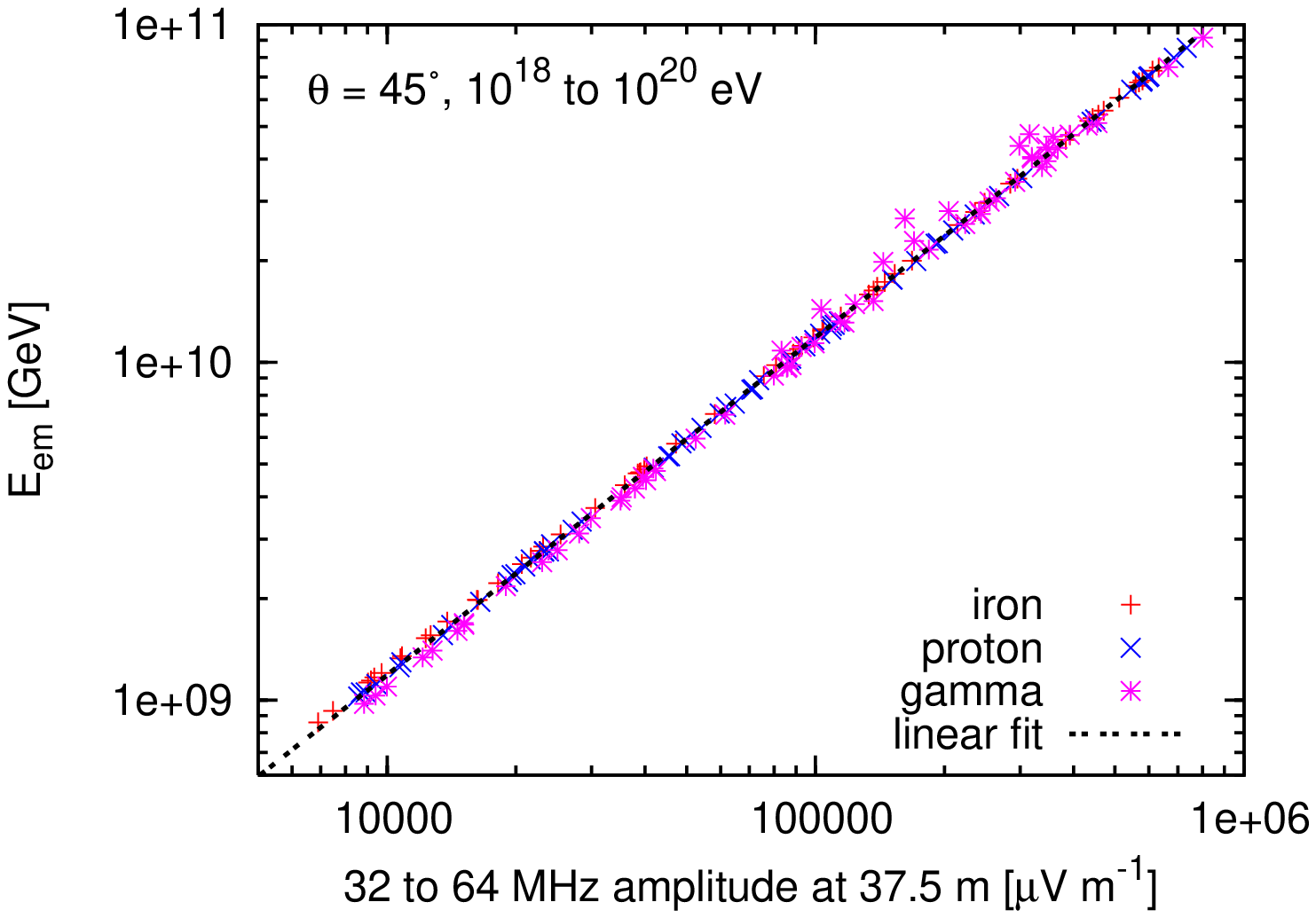}
\caption{\label{energy45deg}A measurement in the {\em flat region} at \unit[37.5]{m} north directly yields the shower $E_{\mathrm{em}}$ of 45$^{\circ}$ zenith angle showers. The fit parameter is $a=118700$.\newline\newline\newline}
\end{minipage} \hspace{1.5pc}
\begin{minipage}{15.5pc}
\includegraphics[angle=270,width=15.5pc]{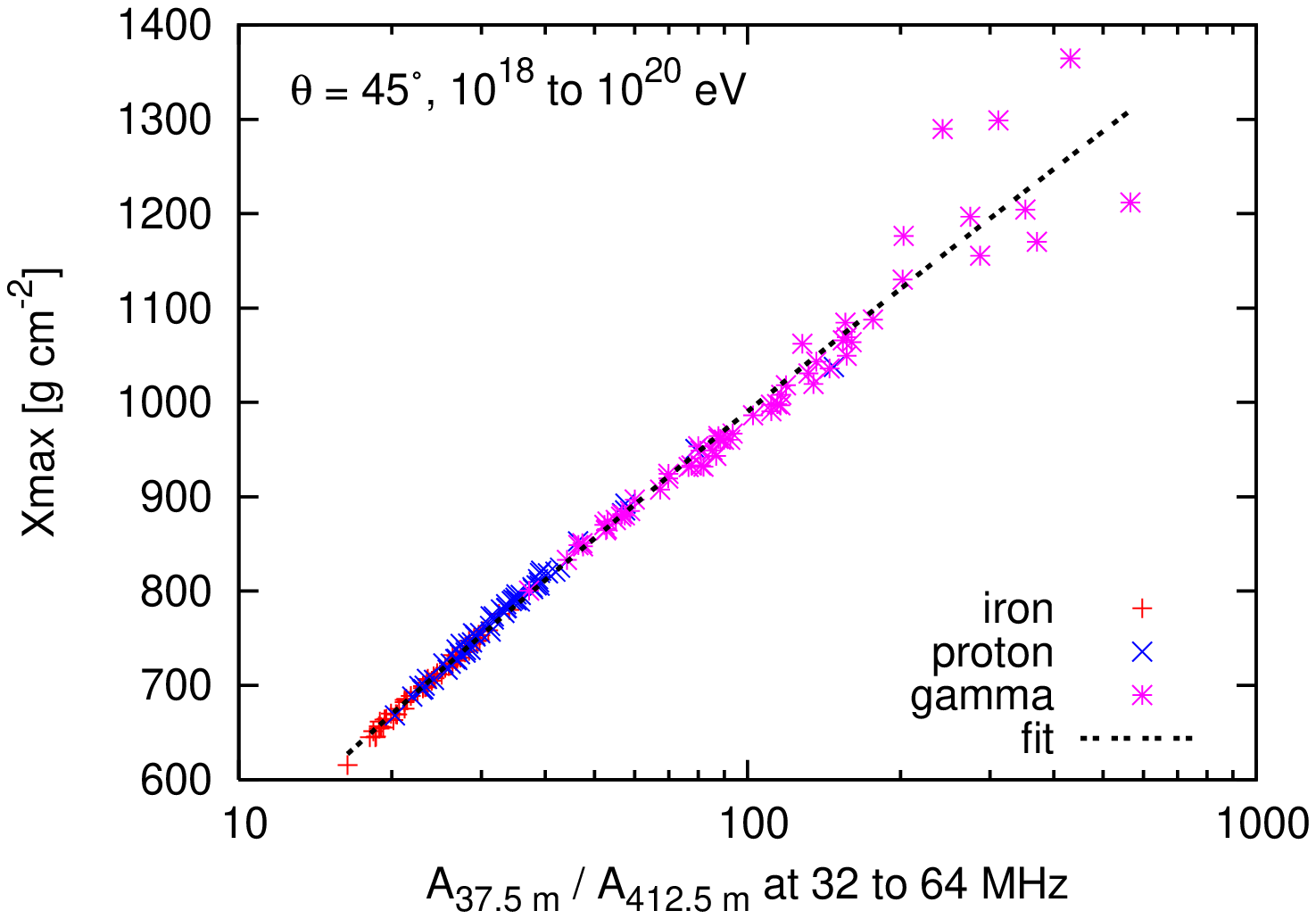}
\caption{\label{composition45deg} The ratio of peak field strengths in the {\em flat region} and {\em steep region} yields the shower $X_{\mathrm{max}}$. The fit parameters for the 45$^{\circ}$ zenith angle showers are $a=312.3$, $b=0.6569$ and $c=0.8062$. The RMS spread amounts to $\sim 19.3\,$g$\,$cm$^{-2}$}
\end{minipage}
\end{figure}

Again, $E_{\mathrm{em}}$ can be determined from a measurement in the {\em flat region} (Fig.~\ref{energy45deg}) and $X_{\mathrm{max}}$ can be deduced from a combined measurement in the {\em flat region} and a {\em steep region} (Fig.\ \ref{composition45deg}). The outliers at the extreme values of $X_{\mathrm{max}}$ are again caused by the LPM effect in high-energy gamma-ray showers, being more prominent in the 45$^{\circ}$ zenith angle case than for 60$^{\circ}$ zenith angle showers.

\section{Conclusions}

Based on Monte Carlo simulations performed with CORSIKA and REAS2, we have shown that geosynchrotron radio emission from air showers of cosmic rays is sensitive to the primary particle energy and mass, and that this information content can be extracted from radio data on a shower-to-shower basis.

For a given observing frequency and zenith angle, there is a well-defined lateral distance region in which the filtered radio peak amplitude is directly proportional to the energy deposited in the atmosphere by the electromagnetic cascade of an air shower, $E_{\mathrm{em}}$. Once the proportionality constant is known, $E_{\mathrm{em}}$ can be derived from a radio measurement in this so-called {\em flat region}. The RMS spread of the $E_{\mathrm{em}}$ values due to shower-to-shower fluctuations amounts to less than 3\% in case of an ideal radio signal measurement. Relating $E_{\mathrm{em}}$ to the energy of the primary particle introduces an additional uncertainty of $\sim 5$ to 7\%.

The peak field strength in the {\em flat region} can be combined with a field strength measured in a {\em steep region}, found at a larger lateral distance and/or higher observing frequency. The ratio of these field strengths can then be directly related to the atmospheric depth of the air shower maximum, $X_{\mathrm{max}}$, on a shower-to-shower basis. The RMS spread of the $X_{\mathrm{max}}$ values around the corresponding fit function due to shower-to-shower fluctuations amounts to $\sim 15$--$20\,$g$\,$cm$^{-2}$, not including any instrumental resolutions.

Exploitation of these signal characteristics with an experimental measurement requires antenna spacings dense enough to allow reconstruction of the electric field strengths in the {\em flat region}. As the location of the {\em flat region} shifts to larger lateral distances for increasing zenith angles, application of this strategy is particularly interesting for inclined showers with at least 45$^{\circ}$, better 60$^{\circ}$, zenith angle. (At smaller zenith angles, the {\em flat} region will shift to distances very close to the shower core, and an extrapolation of the electric field strength to the core region will become necessary.) In the case of 60$^{\circ}$ zenith angle and for an observing frequency of \unit[32--64]{MHz}, the {\em flat region} lies at a (ground-coordinate) distance of $\sim 175\,$m in the azimuthal direction defined by the shower axis. Decreasing the observing frequency, if technically possible, increases this distance considerably. In addition, observations at lower heights above sea level increase the distance, especially for zenith angles smaller than 60$^{\circ}$.

Assuming that the geosynchrotron model accounts for the dominant part of radio emission from extensive air showers, exploitation of these characteristics could thus make radio detection a powerful tool for the determination of the primary energy of cosmic rays and the depth of shower maximum of air showers, which in turn is related to the mass of the primary particle.

%A change of the hadronic interaction models does not change the qualitative behaviour of the radio emission.

\begin{ack}
The authors would like to thank T.\ Pierog and D.\ Heck for their outstanding support regarding many CORSIKA-related aspects of this work, S.\ Laf\`ebre for his efforts in simulating CORSIKA showers on the high-performance computer cluster ``Stella'' and A.\ Haungs and H.\ Falcke for very helpful discussions regarding the analysis and the manuscript. This research has been supported by grant number VH-NG-413 of the Helmholtz Association.
\end{ack}

% \appendix

% \section{}
% \label{}

%________________________________________________________________________

% Bibliographic references with the natbib package:
% Parenthetical: \citep{Bai92} produces (Bailyn 1992).
% Textual: \citet{Bai95} produces Bailyn et al. (1995).
% An affix and part of a reference:
%   \citep[e.g.][Ch. 2]{Bar76}
%   produces (e.g. Barnes et al. 1976, Ch. 2).

%\bibliography{references}

\bibliographystyle{elsart-num}

\end{document}